\input harvmac

\input epsf.tex
\def\figin{\epsfcheck\figin}\def\figins{\epsfcheck\figins}
\def\epsfcheck{\ifx\epsfbox\UnDeFiSIeD
\message{(NO epsf.tex, FIGURES WILL BE IGNORED)}
\gdef\figin##1{\vskip2in}\gdef\figins##1{\hskip.5in}
\else\message{(FIGURES WILL BE INCLUDED)}%
\gdef\figin##1{##1}\gdef\figins##1{##1}\fi}
\def\DefWarn#1{}
\def\figinsert{\goodbreak\midinsert}
\def\ifig#1#2#3{\DefWarn#1\xdef#1{fig.~\the\figno}
\writedef{#1\leftbracket fig.\noexpand~\the\figno}%
\figinsert\figin{\centerline{#3}}\medskip\centerline{\vbox{\baselineskip12pt
\advance\hsize by -1truein\noindent\footnotefont{\bf
Fig.~\the\figno:} #2}}
\bigskip\endinsert\global\advance\figno by1}

\def\K3{{\bf K3}}
\def\journal#1&#2(#3){\unskip, \sl #1\ \bf #2 \rm(19#3) }
\def\andjournal#1&#2(#3){\sl #1~\bf #2 \rm (19#3) }

\def\bar{\overline}
\def\hat{\widehat}

\def\tilde{\widetilde}

\def\frac#1#2{{#1\over#2}}

\def\half{\frac12}

\def\inbar{\,\vrule height1.5ex width.4pt depth0pt}
\def\IC{\relax\hbox{$\inbar\kern-.3em{\rm C}$}}
\def\IR{\relax{\rm I\kern-.18em R}}
\def\IP{\relax{\rm I\kern-.18em P}}

%
%

%
\catcode`\@=11
\def\slash#1{\mathord{\mathpalette\c@ncel{#1}}}
\overfullrule=0pt

\def\EE{{\cal E}}

\def\underrel#1\over#2{\mathrel{\mathop{\kern\z@#1}\limits_{#2}}}

\catcode`\@=12


%

\def \sinh{{\rm sinh}}

\def\exp{{\rm exp}}


\def\hatq{ {\hat q} }
\def\EE{ \epsilon }


\lref\StromingerTM{
  A.~Strominger,
  ``A matrix model for AdS(2),''
  JHEP {\bf 0403}, 066 (2004)
  [arXiv:hep-th/0312194].
}

\lref\WittenIM{
  E.~Witten,
  ``Bound states of strings and p-branes,''
  Nucl.\ Phys.\ B {\bf 460}, 335 (1996)
  [arXiv:hep-th/9510135].
}

\lref\PolchinskiSM{
  J.~Polchinski and A.~Strominger,
  ``New Vacua for Type II String Theory,''
  Phys.\ Lett.\ B {\bf 388}, 736 (1996)
  [arXiv:hep-th/9510227].
}

\lref\OSV{
  H.~Ooguri, A.~Strominger and C.~Vafa,
  ``Black hole attractors and the topological string,''
  Phys.\ Rev.\ D {\bf 70}, 106007 (2004)
  [arXiv:hep-th/0405146].
}

\lref\HananyIE{
  A.~Hanany and E.~Witten,
  ``Type IIB superstrings, BPS monopoles, and three-dimensional gauge
  dynamics,''
  Nucl.\ Phys.\ B {\bf 492}, 152 (1997)
  [arXiv:hep-th/9611230].
}

\lref\KapustinHI{
  A.~Kapustin,
  ``Noncritical superstrings in a Ramond-Ramond background,''
  JHEP {\bf 0406}, 024 (2004)
  [arXiv:hep-th/0308119].
}

\lref\DeWolfeQF{
  O.~DeWolfe, R.~Roiban, M.~Spradlin, A.~Volovich and J.~Walcher,
  ``On the S-matrix of type 0 string theory,''
  JHEP {\bf 0311}, 012 (2003)
  [arXiv:hep-th/0309148].
}

\lref\BerkovitsTG{
  N.~Berkovits, S.~Gukov and B.~C.~Vallilo,
  ``Superstrings in 2D backgrounds with R-R flux and new extremal black
  Nucl.\ Phys.\ B {\bf 614}, 195 (2001)
  [arXiv:hep-th/0107140].
}

\lref\KlebanovWG{
  I.~R.~Klebanov, J.~Maldacena and N.~Seiberg,
  ``Unitary and complex matrix models as 1-d type 0 strings,''
  Commun.\ Math.\ Phys.\  {\bf 252}, 275 (2004)
  [arXiv:hep-th/0309168].
}

\lref\GrossZZ{
  D.~J.~Gross and J.~Walcher,
  ``Non-perturbative RR potentials in the c(hat) = 1 matrix model,''
  JHEP {\bf 0406}, 043 (2004)
  [arXiv:hep-th/0312021].
}

\lref\SeibergNM{
  N.~Seiberg and D.~Shih,
  ``Branes, rings and matrix models in minimal (super)string theory,''
  JHEP {\bf 0402}, 021 (2004)
  [arXiv:hep-th/0312170].
}

\lref\StromingerTM{
  A.~Strominger,
  ``A matrix model for AdS(2),''
  JHEP {\bf 0403}, 066 (2004)
  [arXiv:hep-th/0312194].
}

\lref\GaiottoGD{
  D.~Gaiotto,
  ``Long strings condensation and FZZT branes,''
  arXiv:hep-th/0503215.
}

 \lref\FriessTQ{
  J.~J.~Friess and H.~Verlinde,
  ``Hawking effect in 2-D string theory,''
  arXiv:hep-th/0411100.
}

\lref\DanielssonYI{
  U.~H.~Danielsson,
  ``A matrix model black hole: Act II,''
  JHEP {\bf 0402}, 067 (2004)
  [arXiv:hep-th/0312203].
}

\lref\GukovYP{
  S.~Gukov, T.~Takayanagi and N.~Toumbas,
  ``Flux backgrounds in 2D string theory,''
  JHEP {\bf 0403}, 017 (2004)
  [arXiv:hep-th/0312208].
}

\lref\DavisXB{
  J.~L.~Davis, L.~A.~Pando Zayas and D.~Vaman,
  ``On black hole thermodynamics of 2-D type 0A,''
  JHEP {\bf 0403}, 007 (2004)
  [arXiv:hep-th/0402152].
}

\lref\DanielssonXF{
  U.~H.~Danielsson, J.~P.~Gregory, M.~E.~Olsson, P.~Rajan and M.~Vonk,
  ``Type 0A 2D black hole thermodynamics and the deformed matrix model,''
  JHEP {\bf 0404}, 065 (2004)
  [arXiv:hep-th/0402192].
}

\lref\TakayanagiJZ{
  T.~Takayanagi,
  ``Notes on D-branes in 2D type 0 string theory,''
  JHEP {\bf 0405}, 063 (2004)
  [arXiv:hep-th/0402196].
}

\lref\VerlindeGT{
  H.~Verlinde,
  ``Superstrings on AdS(2) and superconformal matrix quantum mechanics,''
  arXiv:hep-th/0403024.
}

\lref\ParkYC{
  J.~Park and T.~Suyama,
  ``Type 0A matrix model of black hole, integrability and holography,''
  Phys.\ Rev.\ D {\bf 71}, 086002 (2005)
  [arXiv:hep-th/0411006].
}

\lref\SeibergEI{
  N.~Seiberg and D.~Shih,
  ``Flux vacua and branes of the minimal superstring,''
  JHEP {\bf 0501}, 055 (2005)
  [arXiv:hep-th/0412315].
}

\lref\SeibergBX{
  N.~Seiberg,
  ``Observations on the moduli space of two dimensional string theory,''
  JHEP {\bf 0503}, 010 (2005)
  [arXiv:hep-th/0502156].
}

\lref\nonsinglets{
  J.~Maldacena,
   ``Long strings in two dimensional string theory and non-singlets in the
   matrix model,''
  arXiv:hep-th/0503112.
}
\lref\DouglasUP{
  M.~R.~Douglas, I.~R.~Klebanov, D.~Kutasov, J.~Maldacena, E.~Martinec and N.~Seiberg,
  ``A new hat for the c = 1 matrix model,''
  arXiv:hep-th/0307195.
}

\lref\KlebanovWG{
  I.~R.~Klebanov, J.~Maldacena and N.~Seiberg,
  ``Unitary and complex matrix models as 1-d type 0 strings,''
  Commun.\ Math.\ Phys.\  {\bf 252}, 275 (2004)
  [arXiv:hep-th/0309168].
}

\lref\wittenbaryon{
  E.~Witten,
  ``Baryons and branes in anti de Sitter space,''
  JHEP {\bf 9807}, 006 (1998)
  [arXiv:hep-th/9805112].
}

\lref\MorrisBW{ T.~R.~Morris, ``2-D Quantum Gravity, Multicritical
Matter And Complex Matrices,''
FERMILAB-PUB-90-136-T
}

\lref\LapanQZ{
  J.~M.~Lapan and W.~Li,
  ``Falling D0-branes in 2D superstring theory,''
  arXiv:hep-th/0501054.
}

\lref\DavisXI{
  J.~L.~Davis and R.~McNees,
  `Boundary counterterms and the thermodynamics of 2-D black holes,''
  arXiv:hep-th/0411121.
}

%

\lref\MartinecQT{
  E.~Martinec and K.~Okuyama,
  ``Scattered results in 2D string theory,''
  JHEP {\bf 0410}, 065 (2004)
  [arXiv:hep-th/0407136].
}

 \lref\HoQP{
  P.~M.~Ho,
  ``Isometry of AdS(2) and the c = 1 matrix model,''
  JHEP {\bf 0405}, 008 (2004)
  [arXiv:hep-th/0401167].
}

\lref\GrossHE{ D.~J.~Gross and E.~Witten, ``Possible Third Order
Phase Transition In The Large N Lattice Gauge Theory,'' Phys.\
Rev.\ D {\bf 21}, 446 (1980).
}

\lref\DiFrancescoRU{ P.~Di Francesco, ``Rectangular Matrix Models
and Combinatorics of Colored Graphs,'' Nucl.\ Phys.\ B {\bf 648},
461 (2003) [arXiv:cond-mat/0208037].
}

\lref\MartinecHT{ E.~J.~Martinec, G.~W.~Moore and N.~Seiberg,
``Boundary operators in 2-D gravity,'' Phys.\ Lett.\ B {\bf 263},
190 (1991).
}

\lref\minahan{ J.~A.~Minahan, ``Matrix models with boundary terms
and the generalized Painleve II equation,'' Phys.\ Lett.\ B {\bf
268}, 29 (1991);
J.~A.~Minahan, ``Schwinger-Dyson equations for unitary matrix
models with boundaries,'' Phys.\ Lett.\ B {\bf 265}, 382 (1991).
}

\lref\PeriwalGF{ V.~Periwal and D.~Shevitz, ``Unitary Matrix
Models As Exactly Solvable String Theories,'' Phys.\ Rev.\ Lett.\
{\bf 64}, 1326 (1990);
V.~Periwal and D.~Shevitz, ``Exactly Solvable Unitary Matrix
Models: Multicritical Potentials And Correlations,'' Nucl.\ Phys.\
B {\bf 344}, 731 (1990).
}

\lref\DouglasUP{ M.~R.~Douglas, I.~R.~Klebanov, D.~Kutasov,
J.~Maldacena, E.~Martinec and N.~Seiberg, ``A new hat for the c =
1 matrix model,'' arXiv:hep-th/0307195.
}
\lref\AlexandrovFH{
  S.~Y.~Alexandrov, V.~A.~Kazakov and I.~K.~Kostov,
  ``Time-dependent backgrounds of 2D string theory,''
  Nucl.\ Phys.\ B {\bf 640}, 119 (2002)
  [arXiv:hep-th/0205079].
}

\lref\KostovTK{
  I.~K.~Kostov,
  ``Integrable flows in c = 1 string theory,''
  J.\ Phys.\ A {\bf 36}, 3153 (2003)
  [Annales Henri Poincare {\bf 4}, S825 (2003)]
  [arXiv:hep-th/0208034].
}

\lref\AlexandrovPZ{
  S.~Y.~Alexandrov and V.~A.~Kazakov,
  ``Thermodynamics of 2D string theory,''
  JHEP {\bf 0301}, 078 (2003)
  [arXiv:hep-th/0210251].
}

\lref\AlexandrovQK{
  S.~Y.~Alexandrov, V.~A.~Kazakov and I.~K.~Kostov,
  ``2D string theory as normal matrix model,''
  Nucl.\ Phys.\ B {\bf 667}, 90 (2003)
  [arXiv:hep-th/0302106].
}

\lref\YinIV{
  X.~Yin,
  ``Matrix models, integrable structures, and T-duality of type 0
string
  theory,''
  Nucl.\ Phys.\ B {\bf 714}, 137 (2005)
  [arXiv:hep-th/0312236].
}

\lref\AlexandrovKS{
  S.~Alexandrov,
  ``D-branes and complex curves in c=1 string theory,''
  JHEP {\bf 0405}, 025 (2004)
  [arXiv:hep-th/0403116].
}

\lref\AlexandrovCG{
  S.~Y.~Alexandrov and I.~K.~Kostov,
  ``Time-dependent backgrounds of 2D string theory:
Non-perturbative effects,''
  JHEP {\bf 0502}, 023 (2005)
  [arXiv:hep-th/0412223].
}

\lref\TeschnerRD{
  J.~Teschner,
  ``On Tachyon condensation and open-closed duality in the c = 1
string
  theory,''
  arXiv:hep-th/0504043.
}

\lref\CrnkovicMS{ C.~Crnkovic, M.~R.~Douglas and G.~W.~Moore,
``Physical Solutions For Unitary Matrix Models,'' Nucl.\ Phys.\ B
{\bf 360}, 507 (1991).
}

\lref\DiFrancescoXZ{ P.~Di Francesco, H.~Saleur and J.~B.~Zuber,
``Generalized Coulomb Gas Formalism For Two-Dimensional Critical
Models Based On SU(2) Coset Construction,'' Nucl.\ Phys.\ B {\bf
300}, 393 (1988).
}

\lref\DijkgraafPP{ R.~Dijkgraaf, S.~Gukov, V.~A.~Kazakov and
C.~Vafa, ``Perturbative analysis of gauged matrix models,''
arXiv:hep-th/0210238.
}

\lref\AganagicQJ{
  M.~Aganagic, R.~Dijkgraaf, A.~Klemm, M.~Marino and C.~Vafa,
  ``Topological strings and integrable hierarchies,''
  arXiv:hep-th/0312085.
}

\lref\Fukuda{ T.~Fukuda and K.~Hosomichi, ``Super Liouville theory
with boundary,'' Nucl.\ Phys.\ B {\bf 635}, 215 (2002)
[arXiv:hep-th/0202032].
}

\lref\KostovXW{ I.~K.~Kostov, ``Solvable statistical models on a
random lattice,'' Nucl.\ Phys.\ Proc.\ Suppl.\  {\bf 45A}, 13
(1996) [arXiv:hep-th/9509124].
}

\lref\Hollowood{ T.~J.~Hollowood, L.~Miramontes, A.~Pasquinucci
and C.~Nappi, ``Hermitian versus anti-Hermitian one matrix models
and their hierarchies,'' Nucl.\ Phys.\ B {\bf 373}, 247 (1992)
[arXiv:hep-th/9109046].
}

\lref\NappiBI{ C.~R.~Nappi, ``Painleve-II And Odd Polynomials,''
Mod.\ Phys.\ Lett.\ A {\bf 5}, 2773 (1990).
}

\lref\MartinecKA{ E.~J.~Martinec, ``The annular report on
non-critical string theory,'' arXiv:hep-th/0305148.
}

\lref\AlexandrovNN{ S.~Y.~Alexandrov, V.~A.~Kazakov and
D.~Kutasov, ``Non-Perturbative Effects in Matrix Models and
D-branes,'' arXiv:hep-th/0306177.
}

\lref\DalleyBR{ S.~Dalley, C.~V.~Johnson, T.~R.~Morris and
A.~Watterstam, ``Unitary matrix models and 2-D quantum gravity,''
Mod.\ Phys.\ Lett.\ A {\bf 7}, 2753 (1992) [arXiv:hep-th/9206060].
}

\lref\johnsonflows{ C.~V.~Johnson, T.~R.~Morris and A.~Watterstam,
``Global KdV flows and stable 2-D quantum gravity,'' Phys.\ Lett.\
B {\bf 291}, 11 (1992) [arXiv:hep-th/9205056].
}

\lref\DalleyQG{ S.~Dalley, C.~V.~Johnson and T.~Morris,
``Multicritical complex matrix models and nonperturbative 2-D
quantum gravity,'' Nucl.\ Phys.\ B {\bf 368}, 625 (1992).
}

\lref\LafranceWY{ R.~Lafrance and R.~C.~Myers, ``Flows For
Rectangular Matrix Models,'' Mod.\ Phys.\ Lett.\ A {\bf 9}, 101
(1994) [arXiv:hep-th/9308113].
}

\lref\SeibergEB{ N.~Seiberg, ``Notes On Quantum Liouville Theory
And Quantum Gravity,'' Prog.\ Theor.\ Phys.\ Suppl.\  {\bf 102},
319 (1990).
}

\lref\gelfand{ I.~M.~Gelfand and L.~A.~Dikii, ``Asymptotic
Behavior Of The Resolvent Of Sturm-Liouville Equations And The
Algebra Of The Korteweg-De Vries Equations,'' Russ.\ Math.\
Surveys {\bf 30}, 77 (1975) [Usp.\ Mat.\ Nauk {\bf 30}, 67
(1975)].
}

\lref\BrowerMN{ R.~C.~Brower, N.~Deo, S.~Jain and C.~I.~Tan,
``Symmetry breaking in the double well Hermitian matrix models,''
Nucl.\ Phys.\ B {\bf 405}, 166 (1993) [arXiv:hep-th/9212127].
}

\lref\CrnkovicWD{ C.~Crnkovic, M.~R.~Douglas and G.~W.~Moore,
``Loop equations and the topological phase of multi-cut matrix
models,'' Int.\ J.\ Mod.\ Phys.\ A {\bf 7}, 7693 (1992)
[arXiv:hep-th/9108014].
}

\lref\BerKleb{ M.~Bershadsky and I.~R.~Klebanov, ``Genus One Path
Integral In Two-Dimensional Quantum Gravity,'' Phys.\ Rev.\ Lett.\
{\bf 65}, 3088 (1990).
} \lref\BerKlebnew{ M.~Bershadsky and I.~R.~Klebanov, ``Partition
functions and physical states in two-dimensional quantum gravity
and supergravity,'' Nucl.\ Phys.\ B {\bf 360}, 559 (1991).
} \lref\igor{ I.~R.~Klebanov, ``String theory in two-dimensions,''
arXiv:hep-th/9108019.
}

\lref\ginsparg{ P.~Ginsparg and G.~W.~Moore, ``Lectures On 2-D
Gravity And 2-D String Theory,'' arXiv:hep-th/9304011.
}

\lref\DiFrancescoGinsparg{P.~Di Francesco, P.~Ginsparg and
J.~Zinn-Justin,``2-D Gravity and random matrices,'' Phys.\ Rept.\
{\bf 254}, 1 (1995) [arXiv:hep-th/9306153].
}

\lref\joe{J.~Polchinski, ``What is String Theory?''
arXiv:hep-th/9411028
}

\lref\mgv{ J.~McGreevy and H.~Verlinde, ``Strings from tachyons:
The $c = 1$ matrix reloaded,'' arXiv:hep-th/0304224.
}

\lref\SchomerusVV{ V.~Schomerus, ``Rolling tachyons from Liouville
theory,'' arXiv:hep-th/0306026.
}

\lref\GaiottoYF{ D.~Gaiotto, N.~Itzhaki and L.~Rastelli, ``On the
BCFT description of holes in the c = 1 matrix model,''
arXiv:hep-th/0307221.
}

\lref\GutperleIJ{ M.~Gutperle and P.~Kraus, ``D-brane dynamics in
the c = 1 matrix model,'' arXiv:hep-th/0308047.
}

\lref\KapustinHI{ A.~Kapustin, ``Noncritical superstrings in a
Ramond-Ramond background,'' arXiv:hep-th/0308119.
}

\lref\GiveonWN{ A.~Giveon, A.~Konechny, A.~Pakman and A.~Sever,
``Type 0 strings in a 2-d black hole,'' arXiv:hep-th/0309056.
}

\lref\Harv{ J.L.~Karczmarek and A.~Strominger, ``Matrix
cosmology,'' arXiv:hep-th/0309138.
}

\lref\Kitp{O.~DeWolfe, R.~Roiban, M.~Spradlin, A.~Volovich and
J.~Walcher, ``On the S-matrix of type 0 string theory,''
arXiv:hep-th/0309148.
}

\lref\KlebanovKM{ I.~R.~Klebanov, J.~Maldacena and N.~Seiberg,
``D-brane decay in two-dimensional string theory,''
arXiv:hep-th/0305159.
}

\lref\McGreevyEP{ J.~McGreevy, J.~Teschner and H.~Verlinde,
``Classical and quantum D-branes in 2D string theory,''
arXiv:hep-th/0305194.
}

\lref\TeschnerQK{ J.~Teschner, ``On boundary perturbations in
Liouville theory and brane dynamics in noncritical string
theories,'' arXiv:hep-th/0308140.
}

\lref\zz{ A.~B.~Zamolodchikov and A.~B.~Zamolodchikov, ``Liouville
field theory on a pseudosphere,'' arXiv:hep-th/0101152.
}

\lref\KazakovCH{ V.~A.~Kazakov and A.~A.~Migdal, ``Recent Progress
In The Theory Of Noncritical Strings,'' Nucl.\ Phys.\ B {\bf 311},
171 (1988).
}

\lref\TakayanagiSM{ T.~Takayanagi and N.~Toumbas, ``A Matrix Model
Dual of Type 0B String Theory in Two Dimensions,''
arXiv:hep-th/0307083.
}

\lref\SW{N. Seiberg and E. Witten, unpublished.}

\lref\GrossAY{ D.~J.~Gross and N.~Miljkovic, ``A Nonperturbative
Solution of $D = 1$ String Theory,'' Phys.\ Lett.\ B {\bf 238},
217 (1990);
%
E.~Brezin, V.~A.~Kazakov and A.~B.~Zamolodchikov, ``Scaling
Violation in a Field Theory of Closed Strings in One Physical
Dimension,'' Nucl.\ Phys.\ B {\bf 338}, 673 (1990);
%
P.~Ginsparg and J.~Zinn-Justin, ``2-D Gravity + 1-D Matter,''
Phys.\ Lett.\ B {\bf 240}, 333 (1990).
} \lref\JevickiQN{ A.~Jevicki, ``Developments in 2-d string
theory,'' arXiv:hep-th/9309115.
}

\lref\Sennew{ A.~Sen, ``Open-Closed Duality: Lessons from the
Matrix Model,'' arXiv:hep-th/0308068.
}

\lref\DalleyVR{ S.~Dalley, C.~V.~Johnson and T.~Morris,
``Nonperturbative two-dimensional quantum gravity,'' Nucl.\ Phys.\
B {\bf 368}, 655 (1992).
}

\lref\KutasovSV{ D.~Kutasov and N.~Seiberg, ``Number Of Degrees Of
Freedom, Density Of States And Tachyons In String Theory And
Cft,'' Nucl.\ Phys.\ B {\bf 358}, 600 (1991).
}

\lref\GopakumarKI{ R.~Gopakumar and C.~Vafa,
``On the gauge theory/geometry correspondence,''
Adv.\ Theor.\ Math.\ Phys.\ {\bf 3}, 1415 (1999)
[arXiv:hep-th/9811131].
}

\lref\KlebanovHB{ I.~R.~Klebanov and M.~J.~Strassler,
``Supergravity and a confining gauge theory: Duality cascades and
chiSB-resolution of naked singularities,'' JHEP {\bf 0008}, 052
(2000) [arXiv:hep-th/0007191].
}

\lref\MaldacenaYY{ J.~M.~Maldacena and C.~Nunez, ``Towards the
large N limit of pure N = 1 super Yang Mills,'' Phys.\ Rev.\
Lett.\  {\bf 86}, 588 (2001) [arXiv:hep-th/0008001].
}

\lref\VafaWI{ C.~Vafa, ``Superstrings and topological strings at
large N,'' J.\ Math.\ Phys.\  {\bf 42}, 2798 (2001)
[arXiv:hep-th/0008142].
}

\lref\WittenIG{ E.~Witten, ``On The Structure Of The Topological
Phase Of Two-Dimensional Gravity,'' Nucl.\ Phys.\ B {\bf 340}, 281
(1990).
}

\lref\MartinecHT{ E.~J.~Martinec, G.~W.~Moore and N.~Seiberg,
``Boundary operators in 2-D gravity,'' Phys.\ Lett.\ B {\bf 263},
190 (1991).
}

\lref\CachazoJY{ F.~Cachazo, K.~A.~Intriligator and C.~Vafa, ``A
large N duality via a geometric transition,'' Nucl.\ Phys.\ B {\bf
603}, 3 (2001) [arXiv:hep-th/0103067].
}

\lref\BrezinRB{ E.~Brezin and V.~A.~Kazakov, ``Exactly Solvable
Field Theories Of Closed Strings,'' Phys.\ Lett.\ B {\bf 236}, 144
(1990).
}

\lref\DouglasVE{ M.~R.~Douglas and S.~H.~Shenker, ``Strings In
Less Than One-Dimension,'' Nucl.\ Phys.\ B {\bf 335}, 635 (1990).
}

\lref\GrossVS{ D.~J.~Gross and A.~A.~Migdal, ``Nonperturbative
Two-Dimensional Quantum Gravity,'' Phys.\ Rev.\ Lett.\  {\bf 64},
127 (1990).
}

\lref\fzz{V.~Fateev, A.~B.~Zamolodchikov and A.~B.~Zamolodchikov,
``Boundary Liouville field theory. I: Boundary state and boundary
two-point function,'' arXiv:hep-th/0001012.
}

\lref\teschner{ J.~Teschner, ``Remarks on Liouville theory with
boundary,'' arXiv:hep-th/0009138.
}

\lref\AlexandrovFH{
  S.~Y.~Alexandrov, V.~A.~Kazakov and I.~K.~Kostov,
  ``Time-dependent backgrounds of 2D string theory,''
  Nucl.\ Phys.\ B {\bf 640}, 119 (2002)
  [arXiv:hep-th/0205079].
}

\lref\dgjw{  D.~J.~Gross and J.~Walcher,
  ``Non-perturbative RR potentials in the c(hat) = 1 matrix model,''
  JHEP {\bf 0406}, 043 (2004)
  [arXiv:hep-th/0312021].
}

\lref\bruno{
  B.~C.~da Cunha,
  ``Tachyon effective dynamics and de Sitter vacua,''
  Phys.\ Rev.\ D {\bf 70}, 066002 (2004)
  [arXiv:hep-th/0403217].
}


\Title{ \rightline{hep-th/0506141} }
{\vbox{\centerline{Flux-vacua
in Two Dimensional String Theory}}}
\medskip

\centerline{\it Juan Maldacena and Nathan Seiberg}
\bigskip
\centerline{School of Natural Sciences}
\centerline{Institute for Advanced Study}
\centerline{Einstein Drive, Princeton, NJ 08540}

\smallskip

\vglue .3cm

\bigskip
 \noindent
We analyze the two dimensional type 0 theory with background
RR-fluxes.  Both the 0A and the 0B theory have two distinct fluxes
$q$ and $\tilde q$.  We study these two theories at finite
temperature (compactified on a Euclidean circle of radius $R$) as
a function of the fluxes, the tachyon condensate $\mu$ and the
radius $R$.  Surprisingly, the dependence on $q$, $\tilde q$ and
$\mu$ is rather simple.  The partition function is the absolute
value square of a holomorphic function of $y=|q|+|\tilde q| + i
\sqrt{2\alpha'} \mu$ (up to a simple but interesting correction).
As expected, the 0A and the 0B answers are related by T-duality.
Our answers are derived using the exact matrix models description
of these systems and are interpreted in the low energy spacetime
Lagrangian.

\Date{6/05}

\newsec{Introduction}

The renewed interest in noncritical string theories has originated
from their relevance to current topics in string theory
\refs{\mgv\KlebanovKM\TakayanagiSM-\DouglasUP}, like open/closed
duality, holography and D-branes.  These models are interesting
because they have a complete nonperturbative definition and at the
same time can be analyzed exactly.  As such, they are good
laboratories for subtle nonperturbative questions.  In particular,
this is the only framework where flux vacua -- backgrounds with
RR-fluxes -- can be analyzed exactly.  Issues associated with such
flux vacua have already been discussed both in $\hat c=1$ models
\refs{\TakayanagiSM\DouglasUP\KapustinHI
\DeWolfeQF\GrossZZ\StromingerTM\DanielssonYI\GukovYP
\DavisXB\DanielssonXF\TakayanagiJZ\VerlindeGT\HoQP\ParkYC\DavisXI-\SeibergBX}
and in $\hat c<1$ models \refs{\KlebanovWG\SeibergNM-\SeibergEI}.
However, some of the results of the $\hat c=1$ system appeared
confusing and it has been suggested that the system with RR-flux
is related to black holes. The purpose of this paper is to clarify
some of these confusions.

The $\hat c=1$ model is a two dimensional string theory.  The
target space is parameterized by the time $t$ and the spatial
coordinate $\phi$. The background is not translational invariant;
the system has a linear dilaton which makes the string coupling
space dependent
 \eqn\gsphi{g_s(\phi) = e^{-\phi}}
The $\phi \to + \infty$ asymptotic region is characterized by weak
coupling.  Scattering experiments are performed by sending signals
from this asymptotic region and detecting them as they return.
More precisely, the scattering is to and from null infinities
${\cal J}^\pm$; the incoming modes are functions of $\phi+t$ and
the outgoing modes are functions of $\phi-t$.  It has been assumed
that the system does not have another asymptotic region with $\phi
\to -\infty$; i.e.\ there is no separate scattering to and from
that region. Indeed, the strong coupling region has effectively
finite volume (see, e.g.\ \SeibergEI).

In section 2 we discuss the spacetime picture of the two kinds of
models we study, the 0B and the 0A theories.  We review their
spectra and the leading order terms in the spacetime effective
Lagrangian.  We show that the 0B theory has two kinds of
continuous RR-fluxes, $\nu$ and $\tilde \nu$, and 0A theory has
two kinds of quantized RR-fluxes, $q$ and $\tilde q$.

In addition to these two parameters we can also turn on a
``tachyon'' condensate $\langle T(\phi) \rangle = \mu e^{-\phi}$
and study the physics as a function of the real parameter $\mu$.
The analysis of \DouglasUP\ showed that the physics is smooth as a
function of $\mu$.  Finally, we can also study the system with
Euclidean time which is compactified on a circle of radius $R$.
This corresponds to studying the thermodynamics of the theory with
finite temperature ${1 \over 2\pi R}$.

In section 3 we study the exact 0A theory with its two RR-fluxes
and derive an expression for its partition function ${\cal
Z}_A(\mu,q,\tilde q, R)$ as a function of all variables.  We find
that up to a simple (but interesting) term, the dependence on $q$
and $\tilde q$ is only through $\hatq = |q|+|\tilde q|$.
Furthermore, up to the same simple term, the partition function
factorizes as a holomorphic function of $y=\hatq + i
\sqrt{2\alpha'}\mu$ and its complex conjugate.  We interpret the
dependence on $\hatq = |q|+|\tilde q|$ as a result of the presence
of $|q\tilde q|$ fundamental strings in the system. This is
reminiscent of the factorization involved in topological string
computations, see e.g. \OSV\ and references therein.

The presence of two distinct fluxes in the 0A theory, which couple
differently to the tachyon, has led to speculations about the
existence of extremal black hole solutions when the tachyon is not
excited: $\mu=0$. Indeed the lowest order in $\alpha'$ equations
of motion predict such a solution \BerkovitsTG. Unfortunately it
is not possible to trust the leading order equations in the two
dimensional string theory.  The fact that our matrix model results
depend only on $\hatq = |q|+|\tilde q|$ shows that the physics is
essentially the same as the physics with only one kind of flux
that has been analyzed in \DouglasUP, see also \MartinecQT . Such
analysis does not show any indications of a black hole, namely
there is no entropy and there is no classical absorption. So the
matrix model, as analyzed in this paper is not consistent with an
object that could be called a black hole.

In section 4 we study the exact 0B theory.  We view the partition
function of the noncompact Lorentzian theory ${\cal Z}_B$ as a
{\it transition amplitude} between the past and the future. The
value of this amplitude is complex, but is simpler than expected.
Its phase is given by the real part of
$\Xi\left(-2i\sqrt{2\alpha'}(|\nu|+|\tilde \nu|+\mu)\right) +
\Xi\left(-2i\sqrt{2\alpha'}(|\nu|+|\tilde \nu|-\mu)\right)$
for some function $\Xi$ and $\nu, ~\tilde \nu$ are the
Lorentzian RR fluxes. The finite temperature version of the 0B
theory has quantized fluxes $|q|=-2iR|\nu|$ and $|\tilde q|=-2i R
|\tilde \nu|$. The expression for the finite temperature partition
function is related to that of the 0A theory by the expected
T-duality with the following change in the parameters
  \eqn\tdualityi{ R_B =  { \alpha' \over R_A} ~,~~~~~~~~~~~~~~~~
  \mu_B = {R_A \over \sqrt{ 2 \alpha'} }\mu_A }
with the same $q$ and $\tilde q$.

In Appendix A we review and extend a simple formalism for
describing these systems \AlexandrovFH.  It allows us to simply
compute the transition amplitudes both of the 0B and the 0A theory
and to obtain new insights into the nature of the scattering.

\newsec{Spacetime effective Lagrangian}

In this section we focus on the weak coupling end of the target
space, $\phi \to +\infty$ and study the low energy field theory
there.  Since the string coupling is arbitrarily small, the
dynamics is dominated by classical physics, and the leading
approximation to the effective field theory is valid.  In
particular, the massless modes are reliably found by a weak
coupling worldsheet analysis.  Another simplification in this part
of the target space is that a possible tachyon condensate
 \eqn\tachyoncond{\langle T(\phi) \rangle = \mu e^{-\phi}}
can be neglected there.

\subsec{0B}

We start by analyzing the 0B string theory.  The spectrum consists
of two massless scalars an NS-NS ``tachyon'' $T$ and an RR scalar
$C$.

It is clear from the worldsheet description that the theory has
two discrete ${\bf Z}_2$ symmetries \DouglasUP:
 \item{1.}  The first symmetry acts in the worldsheet description
 as $(-1)^{F_L}$ where $F_L$ is the target space fermion number of
 the worldsheet left movers.  It acts on the spectrum as
 \eqn\chargecon{\eqalign{&T\to T \cr &C\to -C}}
 Since it changes the sign of the RR scalar, it changes the charge
 of D-branes; we will can refer to it as charge conjugation.
 \item{2.} A more subtle symmetry acts in the worldsheet description
 as $(-1)^{f_L}$ where $f_L$ is the left moving worldsheet fermion
 number.   It acts on the spectrum as
 \eqn\sduality{\eqalign{&T\to - T \cr &C_L\to C_L \cr &C_R\to
 -C_R}}
 Here $C_L$ and $C_R$ are the {\it target space} left and right
 moving components of $C$.  Hence the action of this symmetry on
 $C$ is a duality transformation.  By analogy with its higher
 dimensional counterpart, we will refer to it as S-duality.

The invariance under S-duality means that the scalar $C$ is
compact and its radius is the selfdual radius.  Therefore, the
asymptotic theory as $\phi \to +\infty$ has an $SU(2) \times
SU(2)$ symmetry.  This symmetry will be important below.

Since $T$ is odd under the S-duality symmetry, the kinetic term of
$C$ has to be of the form ${1\over 8\pi}f(T) (\partial C)^2$ with
$f(-T) = {1\over f(T)}$.  More detailed worldsheet considerations
show that $f(T)=e^{2T}$ and hence the kinetic term is
 \eqn\Ckinetic{{\CL}_{kinetic}={1\over 8\pi} e^{2T }\left[
 (\partial_t C)^2 - (\partial_\phi C)^2\right]}
Therefore, the coupling of $T$ to $C$ breaks the $SU(2)\times
SU(2)$ symmetry to $U(1) \times U(1)$. In particular, the tachyon
condensate $\langle T(\phi) \rangle = \mu e^{-\phi}$  breaks the
S-duality symmetry; more precisely, the theory with $\mu$ is
related by S-duality to the theory with $-\mu$.

Let us examine the equations of motion which arise from \Ckinetic
 \eqn\Ceom{\partial_t (e^{2T} \partial_t C)-\partial_\phi (e^{2T}
  \partial_\phi C)=0}
For $\phi \to +\infty$ we can neglect $T$ in this expression and
$C$ is simply a free scalar at the selfdual radius.  Of particular
interest to us will be the zero momentum solutions of the
equations of motion
 \eqn\zeromC{ {C \over \sqrt{2} }  \approx 2(\nu \phi + \tilde \nu t)= \nu_{in}
 (\phi+t) + \nu_{out} (\phi-t) }
where we used an approximate sign to remind us that this solution
is valid only for $\phi \to +\infty$. The two integration
constants $\nu$ and $\tilde \nu$, or equivalently $\nu_{in}$ and
$\nu_{out}$ are RR-fluxes. Both of them are odd under the charge
conjugation symmetry \chargecon\ and transform under S-duality
\sduality\ as $\nu \leftrightarrow \tilde \nu$, or equivalently,
$\nu_{out} \to -\nu_{out}$. Since both the $\nu$ and $\tilde \nu$
deformations are non-normalizable as $\phi \to +\infty$ they label
backgrounds which are determined by the behavior at infinity and
they do not fluctuate.

The coupling to $T$ in \Ckinetic\ has important consequences.  If
$T(\phi)$ is nonzero the solution \zeromC\ becomes
 \eqn\zeromCT{ {C\over 2\sqrt 2 }= \tilde \nu t +
 \nu \int e^{-2T(\phi)} d\phi }
(note, as a check that as $\phi \to +\infty$ it goes over to
\zeromC).  Consider the effect of $T(\phi)=\mu e^{-\phi}$ with
positive $\mu$ on \zeromCT.  At the strong coupling end $\phi \to
-\infty$ the second term $\nu \int e^{-2T(\phi)} d\phi $ rapidly
goes to zero, and the corresponding mode is normalizable. (Of
course, it is not normalizable as $\phi \to +\infty$.) Hence it is
a standard background deformation.

This is to be contrasted with the first term $\tilde \nu t$.  The
norm of the small fluctuations which is derived from \Ckinetic\ is
$\int d\phi e^{2T(\phi)} \delta C^2$, and hence it is not
normalizable at $\phi \to -\infty$.  Such a deformation, which is
singular in the interior of the target space, can be present only
if an object is present at its singularity.  In our case, the
relevant object is a D-brane which carries RR-charge. It sources
the RR-flux $\tilde \nu $.

For negative $\mu$ the situation is reversed.  The RR-flux $\tilde
\nu $ is normalizable at $\phi \to -\infty$, and it does not need a
D-brane source.    However, the other flux $\nu \int
e^{-2T(\phi)} d\phi $ needs D-branes at $-\infty$. This exchange
in the behavior of the two fluxes under the change of the sign of
$\mu$ is consistent with the S-duality symmetry.

As we vary $\mu$ from positive to negative values the physics has
to change in a continuous fashion.  This is particularly obvious
in the asymptotic weak coupling end where the effects of nonzero
$\mu$ are negligible.  Therefore, we see here a phenomenon which
has already been observed elsewhere \refs{\GopakumarKI\KlebanovHB
\MaldacenaYY\VafaWI-\CachazoJY, \KlebanovWG\SeibergNM-\SeibergEI},
that RR-flux without D-branes can be continuously transformed to
RR-flux carried by D-branes.

We should clarify the nature of these D-branes at infinity. In the
worldsheet description these are the so called ZZ-branes \zz.
Since they couple to the scalar $C$, the relevant branes are
D-instantons.  This means that our background flux represents a
transition which is mediated by instantons.  For positive $\mu$ we
have $\tilde \nu$ D-instantons per unit time and for negative
$\mu$ we need $ \nu$ such instantons per unit time. Although the
number of such D-instantons is quantized, the numbers per unit
time, $\nu$ or $\tilde \nu$ do not have to be quantized.

Let us examine a background with generic $\nu$ and $\tilde \nu$.
It is easy to calculate the energy momentum tensor of that
background as $\phi \to + \infty$.  It is
 \eqn\backTf{T_{++}={1\over
 4\pi}\nu_{in}^2+...\qquad , \qquad T_{--}={1\over
 4\pi}\nu_{out}^2+... \qquad , \qquad T_{+-}=0+...}
Here the ellipses represent $\phi$ dependent corrections which are
negligible as $\phi \to +\infty$. As far as the asymptotic
Lagrangian \Ckinetic\ is concerned, there is no problem with such
a background.  However, a crucial part of the story is that the
dynamics is such that pulses get reflected from the $\phi = -
\infty $ region. Furthermore the reflection from this region
conserves energy. On the other hand, the incoming energy flux from
${\cal J}^-$ which is   $\int dx^- T_{--}$ is not the same as the
outgoing energy flux  through ${\cal J}^+$ which is   $\int dx^+
T_{++}$. (These integrals are  infinite since we have a constant
flux). Therefore, conservation of energy implies that we should
either send in extra excitations from  the past, or produce extra
excitations in the future. For simplicity we can focus on the
lowest energy excitation by adding excitations on the side that
has the lower flux, so as to match the side with higher flux. The
lowest energy configuration has
 \eqn\lowesten{T_{++}=T_{--} = {1\over
 4\pi} \max(\nu_{in}^2, \nu_{out}^2) +
 ...= {1\over
 4\pi} (|\nu|+|\tilde \nu|)^2 + ...}
Depending on whether $\nu_{out}^2$ is bigger or smaller than
$\nu_{out}^2$, this is achieved by adding waves with $T_{t\phi}=
{1\over \pi} \nu \tilde \nu +...$ either in the past or in the
future .

We are going to be interested in computing the  scattering
amplitude between a state in the past which is characterized by
$\nu_{in}$ and a state in the future which is characterized by
$\nu_{out}$.  A full characterization of the states involves
specifying the state for all the oscillators of the fields $T$ and
$C$. All we are doing in this section is to analyze the asymptotic
region in order to understand which states we can send in and
which states we expect to come out. Below, we will extend this
discussion in the asymptotic region to the full system and will
derive the exact expression for
for the scattering amplitude. We will see that it depends only on
$|\nu|+|\tilde \nu|$.

 Finally, we would like to comment on the 0B
theory on a Euclidean circle of radius $R$.  The analytic
continuation to Euclidean space leads to real $\nu_E=i\nu$ and
$\tilde \nu_E=i\tilde \nu$, where the subscript $E$ denotes that
these are the Euclidean values. Here, in this Euclidean time setup
for positive $\mu$ the parameter $\nu_E$ is proportional to the
number of instantons per unit Euclidean time and therefore the
parameter $q=2\nu_E R$ is quantized (the precise normalization
will be derived below). Similarly, for negative $\mu$ the
parameter $\tilde q=2\tilde \nu_E R$ is quantized.  By continuity
these two parameters are quantized for all $\mu$.

\subsec{0A}

The discussion of the 0A string theory parallels that of the 0B
theory.  In fact, when the 0B theory is analytically continued to
Euclidean time and that coordinate is compactified, it is T-dual
to the 0A theory.

The spectrum of the 0A theory consists of an NS-NS ``tachyon''
$T$, but the RR-scalar $C$ is absent.  It is replaced by two gauge
fields $F_{t\phi}$ and $\tilde F_{t\phi}$.  These gauge fields
have no propagating degrees of freedom, and only their zero
momentum values can change.

Again, the theory has two discrete ${\bf Z}_2$ symmetries:
 \item{1.}  The charge conjugation symmetry which acts on the
 worldsheet theory as $(-1)^{F_L}$ acts on these fields as
 \eqn\chargecona{\eqalign{&T\to T \cr &F_{t\phi}\to
 -F_{t\phi}\cr & \tilde F_{t\phi}\to
 -\tilde F_{t\phi}}}
 \item{2.}  The S-duality symmetry which acts on the worldsheet
 theory as $(-1)^{f_L}$ acts on the fields as
 \eqn\sdualitya{\eqalign{&T\to - T \cr &F_{t\phi}\to
 \tilde F_{t\phi} \cr &\tilde F_{t\phi}\to F_{t\phi}}}

The Lagrangian \Ckinetic\ is replaced by
 \eqn\Fkinetic{{\CL}={\pi \alpha' }\left(e^{2T}
 F_{t\phi}^2+e^{-2T} \tilde F_{t\phi}^2 \right)}
which is invariant under the two symmetries \chargecona\sdualitya.
The asymptotic solution of the equations of motion is $ 2 \pi
\alpha' F_{t\phi}=q$, $ 2 \pi \alpha' \tilde F_{t\phi}=\tilde q$.
Including the $T$ dependent prefactors in \Fkinetic\ the solutions
are
 \eqn\Fbackground{\eqalign{& F_{t\phi}=q e^{-2T} \cr & \tilde
 F_{t\phi}=\tilde q e^{2T}}}
For negative $\mu$ background $ F_{t\phi}$ is singular at $\phi
\to -\infty$ and is generated by D-branes at $\phi \to -\infty$,
while the background $\tilde F_{t\phi}$ is regular and does not
need such branes. For positive $\mu$ the situation is reversed.
These D-branes at infinity carry RR-electric charge. As in the 0B
theory, these are charged ZZ-branes.  However, unlike the
D-instantons of the 0B theory the relevant branes couple to gauge
field one forms, and therefore they are D0-branes.  Hence, $q$ and
$\tilde q$ are quantized.

Consider a background with generic quantized values of $q $ and
$\tilde q$.  By analogy with similar situations in critical string
theory \refs{\WittenIM\PolchinskiSM\HananyIE-\wittenbaryon} we
expect that such a background is possible only if we add to the
system $q\tilde q$ fundamental strings.  This expectation can be
derived by examining the coupling to the two form field $B$.  Such
a field does not have interesting dynamics in two dimensions, but
its equation of motion shows that such strings must be present.
This conclusion can also be derived by starting with the Euclidean
0B theory on a circle. In the 0B theory we had to add energy flux
to compensate the imbalance $T_{t\phi} \sim {1\over \pi} \nu
\tilde \nu$. This translates, after rotation to Euclidean space
and T-duality, to adding $q\tilde q$ strings in the 0A theory.
Note that the sign of $q \tilde q$ is correlated with the orientation
of these strings in the two dimensional target space.

\newsec{0A Matrix model}

In this section we consider the matrix model of the two
dimensional 0A string theory \DouglasUP . This is a gauged matrix
model which contains a complex matrix, $m$, which transforms in
the bifundamental of $U(N)_A\times U(N)_B$. There are two ways of
introducing fluxes. First, we can modify the gauge groups so that
we start with $U(N)_A \times U(M)_B$ with $ q = M-N$ (for
simplicity assume that $q>0$). This leads to $\tilde q=0$,
$q\not=0 $. This corresponds to placing $M$ charged ZZ branes and
$N$ anti-ZZ-branes at $\phi = -\infty$ and then letting the open
string tachyon condense. It is clear from this description that as
long as $\mu$ is below the barrier (in our conventions, $\mu<0$),
we will have $q$ charged ZZ branes left over.  So in this set up
we end up describing the configuration with the flux that is
sourced by D-branes. We expect that these ZZ branes will be stuck
at the strong coupling end since a charged ZZ brane does not have
an open string tachyon. One could consider charged D0 branes that
move in the bulk of the two dimensional space \LapanQZ . We expect
that these D0 branes will exist only in the non-singlet sector of
the matrix model, since the Euclidean boundary states that
represent them contain a non-normalizable open string winding
mode.

 The second way to introduce flux is for
$\tilde q \not =0$, $ q=0$. In this case we set $N=M$ and add to
the matrix model a term of the form
 \eqn\csterm{ S = S_0 + i \tilde q \int (Tr A - Tr B) }
where $A$ and $B$ are the gauge fields for the $U(N)_A$ and
$U(N)_B$ gauge groups respectively. As shown in \DouglasUP\ this
leads to a problem where the eigenvalues move in a complex plane,
all with angular momentum $\tilde q$. This can also be viewed as a
special case of the general problem of coupling the matrix model
to non-singlet representations. In this case we simply have a
singlet representation of $SU(N)_A\times SU(N)_B$ which carries
charge under the relative $U(1)$ (which is generated by the
difference between the generators of $ U(1)_A \subset U(N)_A $ and
of $U(1)_B\subset U(N)_B$). Below the barrier, $\mu<0$, we can
understand the origin of \csterm\ as follows. As we explained
above, the charged ZZ branes source the flux $ F$. In this case
the second flux $\tilde F$ can be excited and leads to a smooth
geometry. If we add ZZ branes the flux $\tilde F$ leads to a
Chern-Simons term on the worldvolume of the ZZ branes that
produces \csterm . This is the same type of coupling that leads to
the usual Chern-Simons terms on D-brane worldvolumes in the ten
dimensional superstring.

Surprisingly, once we reduce the problem to eigenvalues, the
dynamics of these two cases is exactly the same \DouglasUP.  Below
we will slightly qualify this general comment.

We can now study the case with non-zero $q$ and $\tilde q$. The
first naive idea is to consider again a rectangular matrix with $M
= N + q$ and add the Chern-Simons term \csterm . Let us assume for
simplicity that $q>0$.  However, as we now explain, in this case
the path integral vanishes.  The matrix model degree of freedom,
the matrix $m$, is not charged under the the diagonal $U(1)$
generated by the sum of the generators of $U(1)_A $ and $ U(1)_B$.
Our normalization for these $U(1)$s  is such that the fundamental
representation of $SU(N) \subset U(N)$ has charge one (modulo
$N$). On the other hand, the coupling \csterm\ leads to charge $-q
\tilde q $ under diagonal $U(1)$. Since this charge cannot be
cancelled, the path integral vanishes. In other words we cannot
obey the Gauss law for this gauge field.

In order to learn how to deal with this, let us return to the
spacetime picture and understand what happens in string theory
when we start with flux $\tilde q$ and we attempt to put a charged
ZZ brane. There is a coupling $ S = -i \tilde q \int  B $ on the
worldvolume of the ZZ brane, where $B$ is the worldvolume $U(1)$
gauge field. In order to cancel this charge we need to have
$\tilde q $ strings ending on the ZZ brane. If we have $q$ charged
ZZ branes we need to add $q \tilde q$ strings ending on them. A
similar situation has been encountered in various situations in
\refs{\WittenIM\PolchinskiSM\HananyIE-\wittenbaryon}. So the
matrix model that contains both fluxes necessarily involves the
presence of a non-trivial representation of $U(M)_B$ with
$M$-ality $q \tilde q $.

In summary, the matrix model with both fluxes is a $U(N)_A\times U(N+  q)_B$
gauged matrix model
\eqn\actt{
 \int {\cal D}(A,B, m)
 e^{ i \int dt Tr[ (D_0 m)^\dagger D_0 m + { 1 \over 2 \alpha'}  m^\dagger m ] }
 e^{ i \tilde q \int(TrA - TrB) } Tr_{\cal R} P e^{i \int B }
 }
where ${\cal R}$ is a representation of $U(N+  q)_B$ with $q\tilde
q$ $M$-ality\foot{In principle we can also introduce a
representation of $U(N)_A$. In this case the constraint is that
the $M$-ality  of the representation of $U(M)_B$ minus the
$N$-ality of the representation of $U(N)_A$ should be $q\tilde
q$.}. We can now analyze this problem using the general procedure
described in \nonsinglets . We diagonalize the matrix $m$ and we
integrate out the gauge fields. Then we get an effective
hamiltonian of the form\foot{ The apparent differences between
this expression and the one in \nonsinglets\ are due to the fact
that in \nonsinglets\   $\tilde q$ is included  as part of the
$U(1)$ charge of the representation.}
 \eqn\hamiltgen{\eqalign{
 H  = & \left[\sum_{i=1}^N -
 \half { \partial^2 \over \partial \rho_i}  - \half \rho_i^2 +
 \half { \tilde q^2  + q^2 - { 1 \over 4} \over\rho_i^2}  + \right.
 \cr
 + & \left. 2 \sum_{ i <  j \leq N} { \Pi_j^{ ~ i} \Pi_i^{~ j}\over
 (\rho_i^2 - \rho_j^2)} + \sum_{i=1}^N  { 1 \over \rho_i^2}
 \sum_{j>N} (   \Pi_j^{~ i}   \Pi_i^{~ j} +
  \Pi_i^{~j} \Pi_j^{~ i} ) \right] P_0
 }}
where $\Pi_i^j$ are the $U(N+q)$ generators in the representation
${\cal R}$ and $P_0$ is a projector on the states obeying
\eqn\projoa{\eqalign{
 & \Pi_{i}^i  = 0 ~~~~({\rm no~sum})~,~~~~~~~~i=1,\cdots ,N
 \cr
 &  \Pi_l^{~ k} = \tilde q \, \delta_l^k ~,~~~~~l,k>N
 }}
The last condition implies that under the decomposition $U(N+q)\to
U(N) \times U(q)$ we select states that are singlets of $SU(q)$
and carry $U(1)_q\subset U(q)$ charge $q \tilde q$.  The simplest
way in which we can achieve this is by starting out with an
$SU(N+q)$ representation whose Young tableaux contains $ q$ rows
of length $\tilde q$ (we assume that $\tilde q>0$), see figure
1(a). Let us call this representation ${\cal R}_0$. In this case
the state that is a singlet under $SU(q)$ is also a singlet under
$SU(N)$ and obeys the two conditions \projoa .

\medskip \ifig\figI{The Young tableaux (a) corresponds to the
simplest representation ${\cal R}_0$ which leads to a nontrivial
answer.  The Young tableaux (b) is a more complicated
representation which also contributes. The letters and numbers
along the sides of the diagrams denote the number of boxes in that
side.} {\epsfxsize=0.6\hsize\epsfbox{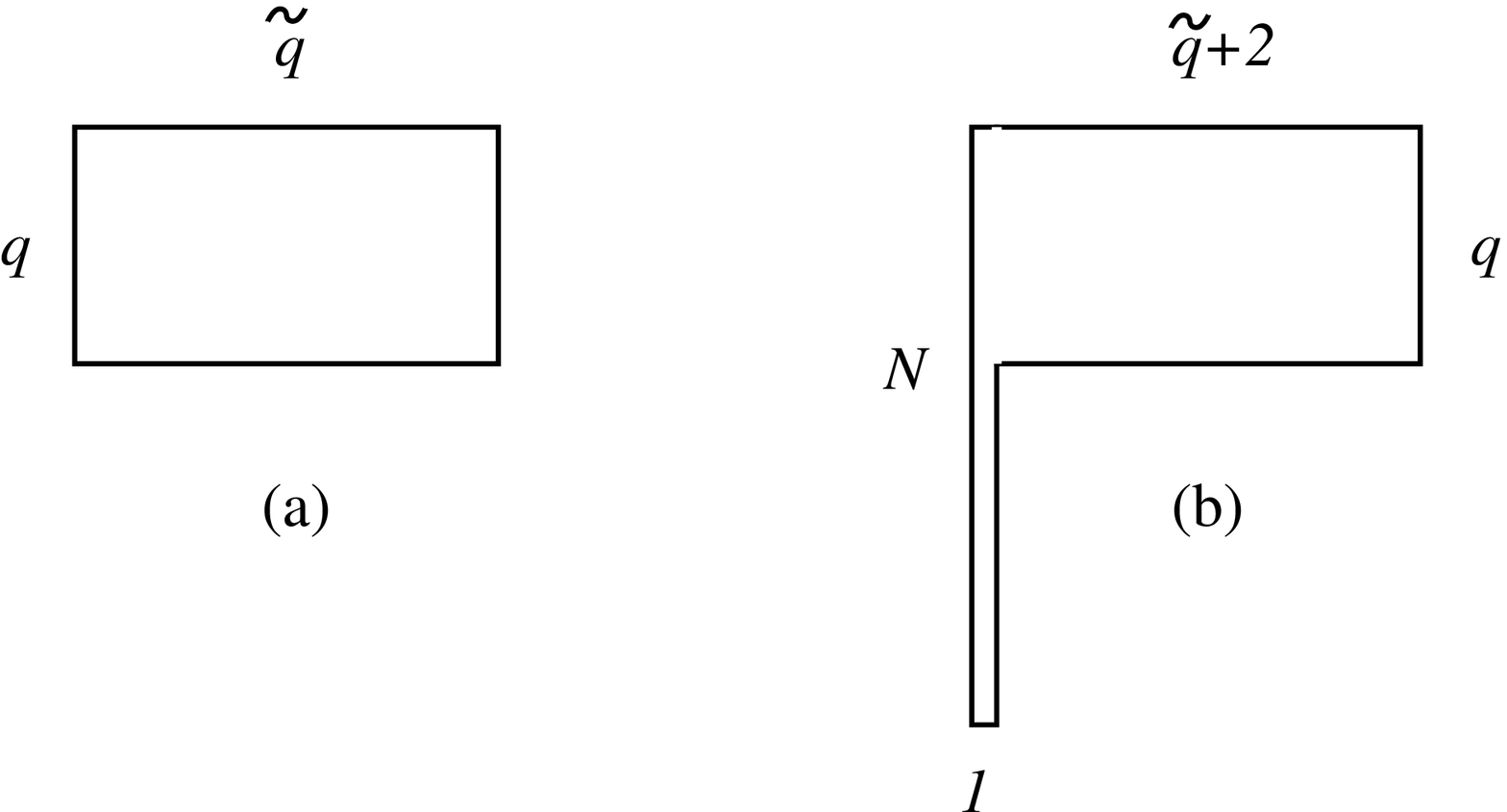}}

We now need to consider the operator that appears in the
Hamiltonian \hamiltgen\
 \eqn\operap{ Q_i =\sum_{j>N} (   \Pi_j^{~
 i} \Pi_i^{~ j} +
  \Pi_i^{~j} \Pi_j^{~ i} )  ~,~~~~~~~~~~i\leq N
 }
This operator transforms in the singlet of $SU(q)$ and it
decomposes as the singlet plus adjoint in $SU(N)$. This implies
that when we act on the single state that is $SU(q)$ invariant in
the representation ${\cal R}_0$ it can give us a state in the
adjoint or the singlet of $SU(N)$. Since the only state in ${\cal
R}_0$ that is in the singlet of $SU(q)$ is also in the singlet of
$SU(N)$, we conclude that the action of $Q_i$ can only give us a
singlet. So this action will just be a c-number. We can simply
compute this c-number to be
 \eqn\eigenv{ Q_i = q \tilde q }
Going back to the hamiltonian we find that it reduces to
 \eqn\simpleeffh{ H = \sum_{i=1}^N -
 \half { \partial^2 \over \partial \rho_i}  - \half \rho_i^2 +
 \half { (\tilde q  + q)^2 - { 1 \over 4} \over
 \rho_i^2}
 }
Here we have assumed that $q$ and $\tilde q$ are positive. The
same analysis can be repeated for the general case and we find
that the dynamics depends only on
  \eqn\qhatd{\hatq = |q| + |\tilde q|}
This is a surprising result from the point of view of the target
space theory, as well as the matrix model.

Before we continue, let us explain what happens if our
representation is a more general representation that contains a
state obeying \projoa . An example of a more general
representation can be found in figure 1(b). In this case a state
that obeys the second condition in \projoa\ can be in the singlet
or the adjoint of $SU(N)$. The operator in \operap\ mixes the
singlet with the adjoint of $SU(N)$. So we expect that all these
states will have properties that are similar to those encountered
in general non-singlet representations as discussed in
\nonsinglets. Such states  have a divergent energy gap compared to
the state that comes from the representation ${\cal R}_0$ in
figure 1(a). In the spacetime theory these states can be
understood as states that, besides the $q \tilde q$ strings ending
on the charged ZZ brane, contain more string anti-string pairs.
The fact that these strings stretch all the way to infinity is
related to this divergence in the energy \nonsinglets . This
divergence implies that these other states are in a different
superselection sector. We would have found similar divergencies,
due to extra strings, if we had also introduced a representation
of the first group $U(N)_A$. So from now on we will assume that we
are adding simply the representation ${\cal R}_0$ in figure 1(a).

Finally, let us present an alternate physical interpretation of
the need for the representation ${\cal R}_0$.  When the two kinds
of branes/fluxes are present the system has massive open string
fermions\foot{These are similar to the fermions in the D0-D8
system, except that here they come from open strings stretched all
the way to infinity, and hence they are infinitely massive.}.
These can be added to the matrix model. Their quantization leads
to several representations which ultimately, after the use of the
Gauss law constraints, lead to ${\cal R}_0$.

The finite temperature partition function can then be computed as
in \DouglasUP\ where the case $\tilde q=0$ and arbitrary $q$ was
studied.  We repeat this computation in Appendix A.  Our arguments
that it is a function of $\hatq = |q| + |\tilde q|$ alow us to
extend it to
 \eqn\zeroaan{ \partial_\mu^3 \log{\cal Z}_{0A} = - Re {1\over 2}
 \int_0^\infty { dt \over t} \partial_\mu^3 e^{-( \hat q + i
 \sqrt{2\alpha'} \mu) {t\over 2} } { 1 \over \sinh {t \over 2}
 \,  \sinh \sqrt{\alpha'\over 2}  { t \over 2R}}
 }
Note that the integral in the right hand side converges. When we
integrate this expression three times with respect to $\mu$ we
need three integration constants -- a $q$ dependent second order
polynomial in $\mu$.  We claim that the answer is
 \eqn\result{ \eqalign{
 &\log {\cal Z}_{0A}(\mu,q,\tilde q, R) =
 \Omega\left(y , r\right)
 + \Omega\left(\bar y, r\right) + ( 2 \pi R) { \mu \over 4}
 (|q| - |\tilde q| ) \cr
 &y=|q|+|\tilde q| + i \sqrt{2 \alpha' } \mu \cr
 &r=R \sqrt{ 2 \over \alpha'}
 }}
where the function $\Omega(y,r)$ is given by
 \eqn\ndwdfr{
 \Omega(y,r)  \equiv - \int_0^\infty { dt \over t}\left[ e^{ -
 {y t \over 2}} { 1 \over  4  \sinh   { t \over 2} \sinh
 { t\over 2r}} - {  r \over t^2}  +  { r y \over 2 t}  +
 [ {1 \over 24} (r + {1\over r}) - {r y^2 \over 8}]e^{-t} \right]
 }
We are interested in $Re(y)=\hatq \ge 0$ where this integral
converges.  (More generally, it converges for $Re(y)> -(1 + { 1
\over r}) $.)  Therefore, $\Omega$ is an analytic function of $y$
which is real along the positive real $y$ axis. It is interesting
that a closely related function appears in a totally different
context in \fzz \foot{ In terms of the function $G(x)$ in \fzz\
our function is $\Omega(y,r) = C(b) + \log G( { y \over 2 b} +
{Q\over 2})   $ where $b = \sqrt{r}$, $Q = b + b^{-1}$ and $C(b)$
is a constant independent of $y$.}.

The expression \result\ for the partition function is one of the
main results of this paper.  Let us discuss it in more detail.

The last term in \result\ depends on $q$ and $\tilde q$ separately
and not only on $\hatq=|q|+|\tilde q|$.  We will return to it
below.

It is surprising that up to this last term in \result\ the
complicated function $\log {\cal Z}_{0A}$ is given as a sum of a
holomorphic and an anti-holomorphic functions. Correspondingly,
the partition function $ {\cal Z}_{0A}$ satisfies {\it holomorphic
factorization}.  The polynomial in $\mu$ which is not determined
by \zeroaan\ was fixed such that this holomorphic factorization is
satisfied.  In the next section we will present another
computation of this partition function where some of this
polynomial dependence on $\mu$ is independently determined.

It is straightforward to work out the asymptotic expansion of
$\Omega(y,r)$ at large $y$ with $Re(y) \ge 0$
 \eqn\asympU{\Omega(y,r) =( \log {y \over 2} - {3 \over 2}){r
 y^2\over 8} - {1\over 24} (r+{1\over r})\log {y \over 2}
 -{7 r^2 + 10 + {7\over r^2}\over 1440}{1\over r y^2}
 + \CO({1\over y^4})}
which leads to the following expression at large $\mu$
 \eqn\freeexp{\eqalign{
 \Omega(y=\hatq+i\sqrt{2\alpha'}\mu ,r)+&\Omega(\bar y=\hat q-
 i\sqrt{2\alpha'}\mu,r)  =
 \left({3 \over 2}- \log (\sqrt{\alpha'\over 2}|\mu|) \right)
 {\alpha' r \mu^2\over 2} -  {2\pi R |\mu| \hatq \over 4}  \cr
 & + \left({\hatq^2 r \over 4}  -{1\over 12}(r+ {1\over r}) \right)
 \log(\sqrt{\alpha'\over 2}|\mu|)\cr
 & +{{7\over r^2} + 10  + 7 r^2+15 \hatq^2 r(\hatq^2
 r- 2 (r+ {1\over r}) )\over 1440 \alpha' r\mu^2} +\CO\left({1\over
 \mu^4}\right)}}
In the worldsheet genus expansion these terms have the following
interpretation.  The first term corresponds to the sphere
contribution. The scaling of the second term which is proportional
to $\hatq$ shows that it arises from a disk diagram. We will
return to this term below. The $\hatq$ independent term in the
second line is the contribution of the torus and the term
proportional to $\hatq^2$ corresponds to a sphere with two RR
insertions, or an annulus. Higher orders can be discussed
similarly.

Let us focus on the term $- 2\pi R |\mu| \hatq/4$ in \freeexp,
which comes from the first term in \asympU . Despite appearance,
because of the absolute value sign on $\mu$, it should not be
discarded as an un-interesting analytic term. We can now
understand the role of the last term in \result. Combining it with
$- 2\pi R |\mu| \hatq/4$ we have
 \eqn\termsl{- {2\pi R|\mu| \over 4}(|q|+|\tilde q|)+ {2\pi R \mu
 \over 4} (|q|-|\tilde q|)=\cases{-\pi R \mu |\tilde q| & $\mu>0$ \cr
 -\pi R |\mu| | q| & $\mu <0$}}
We interpret this as the contribution of the disk amplitude of the
charged ZZ-branes.  For positive $\mu$ we have $|\tilde q|$
ZZ-branes and for negative $\mu$ we have $|q|$ ZZ-branes. Each has
energy $|\mu|/2$. (Recall that the energy of a brane-anti-brane
pair is equal to $|\mu|$. So the energy of a single charged
D-brane should be equal to $|\mu|/2$.). From the point of view of
the 0A matrix model we needed to introduce the last term in
\result\ ``by hand" in order to obey  \termsl . This is an
analytic term in $\mu$, so it is, in principle, possible to
introduce it. However, we will see that this term emerges
naturally from the 0B matrix model.

So after including the analytic term we find precisely the
expected behavior for the free energy. For $\mu <0$ we have
D-branes that produce flux proportional to $ q$ and there are no
terms that have odd powers in $\tilde q$ in the asymptotic
expansion. On the other hand for $\mu>0$ we have the opposite
situation, since now the flux $\tilde q$ is sourced by D-branes.

Despite this simple physical interpretation, our result is still
surprising.  With the exception of the disk term \termsl\ the
semiclassical expansion includes only even powers of $\mu$ and $q$
and $\tilde q$.  This means that there are no contributions from
worldsheets with odd number of boundaries.  For positive $\mu$
this is the expected result when $\tilde q=0$ and there are no
ZZ-branes. Similarly, for negative $\mu$ this is the expected
result when $ q=0$.  The dependence on $q$ and $\tilde q$ through
$\hatq= |q|+|\tilde q|$ together with these expected results
guarantee that, with the exception of the disk \termsl, there are
no contributions from surfaces with odd number of boundaries. We
do not have a worldsheet or spacetime interpretation of this
surprising result.

\newsec{0B Matrix model  }

\subsec{Lorentzian 0B model}

In this section we consider the 0B matrix model which consists of
a hermitian matrix model with an inverted harmonic oscillator
potential such that in the free fermion description we fill the
two sides of the inverted harmonic oscillator potential
\refs{\TakayanagiSM,\DouglasUP}.

To analyze this problem it is useful to realize that the
asymptotic region of the weak coupling end in the target space
geometry is associated to the asymptotic region of the Fermi sea
far away from the maximum of the potential. So the two RR fluxes
$\nu, ~\tilde \nu$ that we discussed in section 2 are associated
with the Fermi levels of the fermions on the two sides of the
potential. Far from the maximum of the potential we can
approximate the fermions as relativistic fermions since the depth
of the Fermi sea is much larger than any finite energy we
consider. It is also possible, and useful, to consider a basis for
the inverted harmonic oscillator problem where the fermions are
exactly relativistic \AlexandrovFH. We review and extend this
formalism in detail in Appendix A. There are actually two possible
bases, which are naturally associated to the coordinates $u
={1\over \sqrt 2}( p -x)$ and $s ={1\over \sqrt 2}( p + x)$. These
are the bases of {\it in} and {\it out} states and the S-matrix
gives the relation between them. This relation is simply a Fourier
transform.

So when we think about the asymptotic states we should think in
terms of relativistic fermions. The asymptotic states live in  the {\it
in} and {\it out} Hilbert spaces of the fermions that are going
towards the maximum of the potential or away from it. Each of these
Hilbert spaces is described by two complex fermions
 \eqn\comle{
 \psi_{\pm}^{in} ~,~~~~~~~\psi^{in \,\dagger \pm } ~;~~~~~~~~~~~~
 \psi_{\pm}^{out} ~,~~~~~~~~~~\psi^{out \, \dagger \pm} }
The $+/-$ indices denote fermions that are moving towards the
right/left. It is very important not to confuse right and left
moving matrix model fermions (denoted here by $+/-$), which are
moving to the right or left in eigenvalue space, with left and
right movers in spacetime, which are related to incoming or
outgoing states\foot{ Also, in the matrix model, do not confuse
right and left {\it moving} fermions with fermions that are to the
left and right side of the potential. For example, the {\it in}
right moving fermion is to the left of the potential.}. Our
notation emphasizes the charge of the fermion under the $U(1)$
current which measures the number of right minus left moving
fermions. This is the current associated to the scalar $C$ in
spacetime. More precisely,
 \eqn\preciserel{\eqalign{
 i (\partial_t + \partial_\phi) C & \, \, \sim \, \,
 \psi^{in \, \dagger +} \psi^{in}_+ - \psi^{in \, \dagger -} \psi^{in}_-
 \cr
 i (\partial_t - \partial_\phi) C & \, \, \sim \, \,
 \psi^{out \, \dagger +} \psi^{out}_+ - \psi^{out \, \dagger -} \psi^{out}_-
 }}
In principle we can specify freely the four Fermi levels of these
four fermions. The fact that fermion number is conserved implies
one relation between these four levels. So we have three
independent levels which denote by $\mu$, $\nu_{in}$ and
$\nu_{out}$. These are defined by saying that $\mu \pm
\nu_{in,out}$ are the Fermi levels associated to the right and
left moving incoming and outgoing fermions (see Figure
2).\foot{When the time is rotated to Euclidean space we must also
rotate $\nu \to i \nu$.  This leads to an imaginary shift of the
Fermi surface.  This is consistent with the analysis of the $\hat
c<1$ systems which are similar to Euclidean $\hat c=1$ where the
RR flux was interpreted as an imaginary shift of the Fermi surface
\refs{\KlebanovWG\SeibergNM-\SeibergEI}.}

\medskip \ifig\figII{ Configurations with generic $\nu_{in,out}$
describe scattering amplitudes.  Figures (a) and (b) describe
scattering below the potential barrier $\mu<0$, and figures (c)
and (d) describe scattering above the potential barrier $\mu>0$.
Figures (a) and (c) describe the initial configurations, while (b)
and (d) describe the final configuration. The dotted line
represents the Fermi level characterized by $\mu$. (Even though we
have represented the Fermi surface reaching all the way to the
potential wall, we really should think of these configurations as
asymptotic states, or as states in the {\it in} or {\it out} basis
defined in the text.)  Note that, as in the figure, the dominant
scattering amplitudes for $\mu>0$ have $\nu_{in} \approx
\nu_{out}$; i.e.\ $\tilde \nu \approx 0$, while for $\mu<0$ they
have $\nu_{in} \approx -\nu_{out}$; i.e.\ $\nu \approx 0$. Note
that in (b) $\nu_{out} <0$.}
{\epsfxsize=0.6\hsize\epsfbox{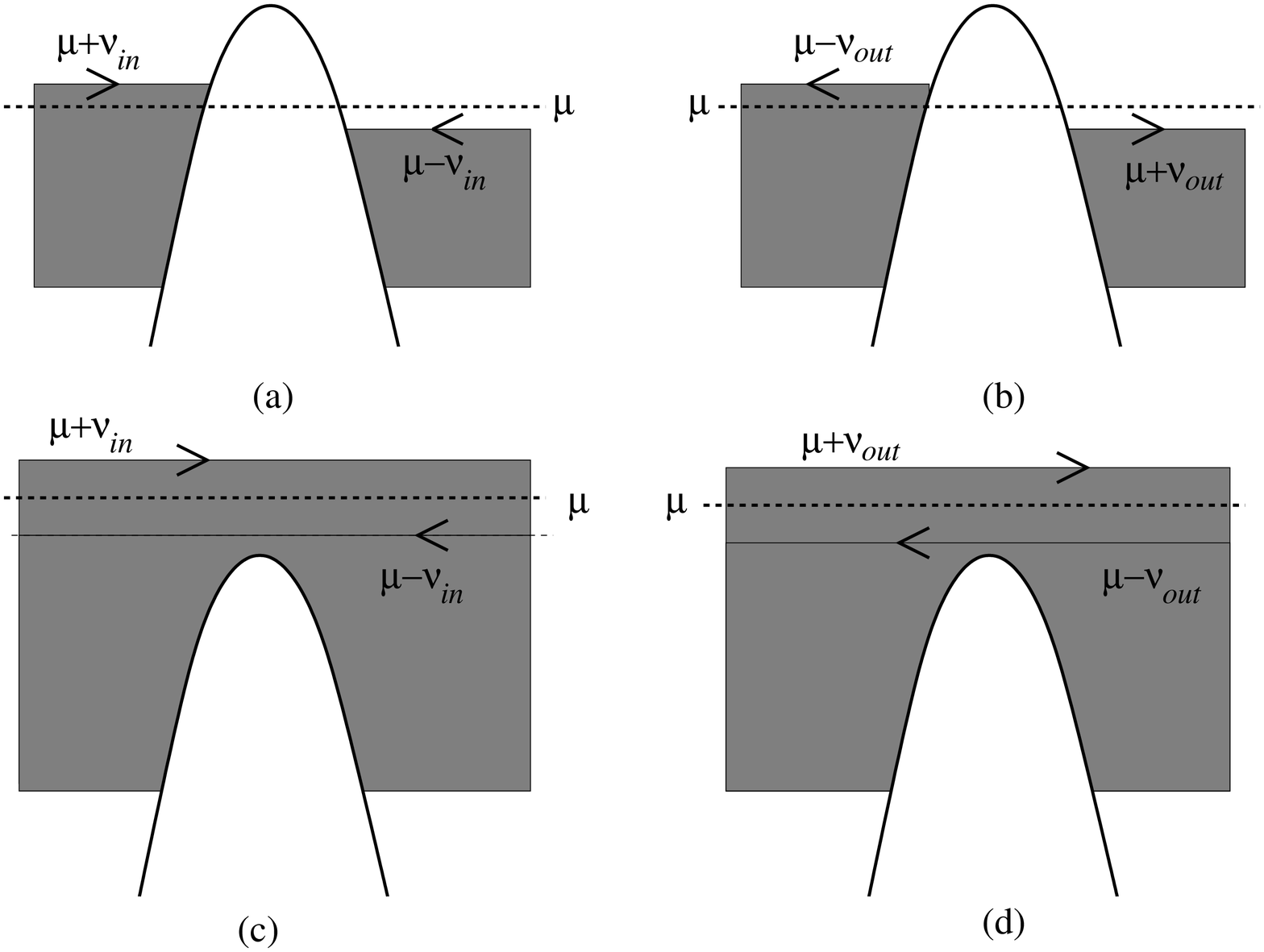}}

\medskip \ifig\figIII{In (a) we see an initial configuration of
incoming fermions with  $\nu_{in}>0$. In (b),(c) we see a outgoing
configurations with $\nu_{out} >0$ and $\nu_{out} <0$
respectively. For the combination (a) (b) we have $\nu_{out} =
\nu_{in}$ and therefore $\tilde \nu =0$, $\nu = \nu_{in}$. On the
other hand for the combination (a), (c) we have $\nu_{out} = -
\nu_{in}$ or $\nu =0$, $\tilde \nu=\nu_{in}$.}
{\epsfxsize=0.6\hsize\epsfbox{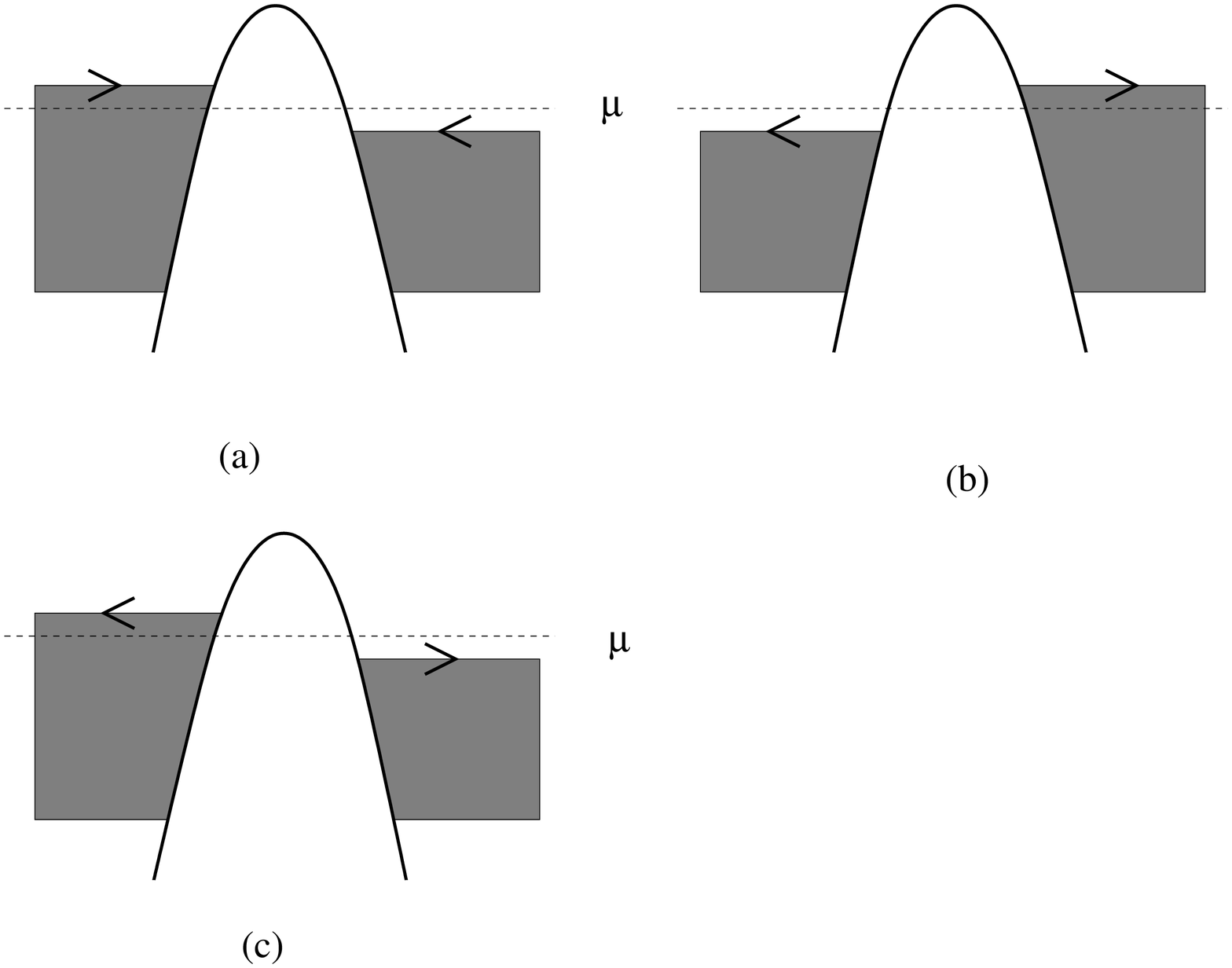}}

In the case that we set $|\nu_{in}| \not = |\nu_{out}|$ we find
that the incoming energy flux is not the same as the outgoing
flux. Therefore we will need to add additional excitations. This
is the same as in the discussion of the spacetime theory in
section 2.

All the information about the inverted harmonic oscillator
potential is contained in the map between in and out states (see
Appendix A)
 \eqn\mapinout{ \psi_{a , \epsilon}^{out}  = \sum_{b=\pm 1}
 {\cal S}_{a}^{\, b} (\epsilon) \psi_{b , \epsilon}^{in} = {
 \Gamma(\half - i \sqrt{2 \alpha'} \epsilon )\over\sqrt{2 \pi}}\sum_{b=\pm 1}
 e^{i {\pi\over 2} ab({1\over 2} -i \sqrt{2 \alpha'} \epsilon )}
  \psi_{b , \epsilon}^{in}
 }
where $a, b = \pm 1$ and $\psi_{a, \epsilon}^{in,out}$ denote the
annihilation operator for a fermion of energy $\epsilon$.

We will now compute the transition amplitude between an {\it in}
state with Fermi levels $\mu$, $\nu_{in}$ to an {\it out} state
with Fermi levels $\mu$, $\nu_{out}$
 \eqn\amplic{ {\cal A} = \langle out(\mu, \nu_{out}) |
 in(\mu,\nu_{in}) \rangle }
and interpret it as the partition function of the 0B theory with
nonzero $\nu$ and $\tilde \nu$: $ {\cal A}={\cal
Z}_{0B}(\mu,\nu,\tilde \nu)$.

For simplicity let us first assume that $\nu_{in} = \nu_{out}
=\nu>0$.  Consider the {\it in} state. The right moving fermions,
which are created by $\psi^{in +}$ are filled up to the Fermi
level $\mu + \nu$, while the left moving fermions, created by
$\psi^{in -}$ are filled up to the Fermi level $\mu - \nu$. The
same is true for the {\it out} fermions. So the overlap is given
by\foot{This expression for $\cal A$ suffers from a phase
ambiguity. We can transform the incoming and outgoing, left and
right moving Hilbert spaces by arbitrary energy independent
phases. These four phases can be used to remove terms linear in
$\mu$, $\nu$ (and later $\tilde \mu$) in $\log{\cal A}$.}
 \eqn\overlap{ {\cal A} = \prod_{ -\infty < \epsilon_n < \mu-\nu }
 [{\cal S} _{+}^{\, +}(\epsilon_n) {\cal S}_{-}^{\, -}(\epsilon_n)
 - {\cal S}_-^{\, +}(\epsilon_n) {\cal
 S}_+^{\, -} (\epsilon_n)] \prod_{ \mu - \nu < \epsilon_m < \mu +
 \nu }
 {\cal S}_{+}^{\, + }(\epsilon_m)
 }
where we have regularized the continuum by putting the system on a
circle of length $L$ so that the density of states is $dn = { L
\over 2 \pi } d \epsilon  $. We have used that up to the energy
$\mu -\nu_{in}$ both states are occupied. Note that the Fermi
statistics produces the determinant of ${\cal S}$ for these
states. On the other hand, for energies in the band between $\mu
-\nu_{in}$ and $\mu + \nu_{in}$ we have only the amplitude for an
incoming right fermion going to an outgoing right fermion. Taking
the logarithm of \overlap\ and expressing the resulting sums in
terms of integrals we obtain\foot{ We set $\alpha'=\half$.}
 \eqn\obtainanb{\eqalign{
 \log{\cal A} =&  { L \over 2\pi}  \left[ \int_{-\Lambda}^{\mu -\nu}
 d\epsilon  \log \left( \Gamma( \half -i \epsilon)/\Gamma(\half
 + i \epsilon) \right) + \right. \cr
 &\qquad + \left. \int_{\mu-\nu}^{\mu + \nu}d\epsilon \log
 (\Gamma( \half - i \epsilon)/\sqrt{2 \pi}) + { \pi\over 2 }
 \int_{\mu-\nu}^{\mu+\nu} d\epsilon\ \epsilon \right] \cr
 =&  { L \over 2\pi}  \left[ \int_{-\Lambda}^{\mu +\nu}
 d\epsilon  \log \left( \Gamma( \half -i \epsilon)\over
 \sqrt{2\pi}\right) - \int_{-\Lambda}^{\mu -\nu} d\epsilon
 \log \left( \Gamma( \half +i \epsilon)\over \sqrt{2\pi}\right)
  + \pi\mu\nu \right]  }}
Here $\Lambda$ is a cutoff on the bottom of the Fermi sea. Using
 \eqn\formla{
 \log\left( \Gamma( \half + z)/\sqrt{2 \pi} \right)  =
 \int_0^\infty { d t \over t} \left[{ e^{ - z t} \over 2 \, \sinh
 {t\over 2} }  - { 1 \over   t} +  z   e^{-t} \right] }
and defining $\Xi(y)$ as related to the large radius limit of
$\Omega(y,r)$ of \ndwdfr
 \eqn\defh{ \Xi(y) \equiv \lim_{r \to \infty}{
 \Omega(y,r) \over 2\pi r} =- {1\over 2\pi} \int_0^\infty
 { dt \over t} \left[ e^{ - y t \over
 2} { 1 \over 2 \, t  \, \sinh {t\over 2} } - { 1 \over t^2} + { y \over
 2 t} + ( { 1 \over 24} - {y^2 \over 8} )e^{-t} \right]
 }
\obtainanb\ becomes
 \eqn\obtainan{
 \log{\cal A}= -i  L  \left[-\Xi ( -2i(\mu + |\nu|)) - \Xi
 (2i(\mu - |\nu|)) +i  \half \mu |\nu| \right]}
where we suppressed a $\Lambda$ dependent imaginary constant and
extended the answer to negative $\nu$ using the charge conjugation
symmetry (left-right symmetry in the matrix model) $\nu \to -\nu$.

Note that the 0B free energy at $\nu=0$  in the non-compact limit
is given by
 \eqn\freeen{ F_{0B} = - \lim_{\beta \to \infty}
 { \log {\cal Z}_{ 0B } \over \beta } = \lim_{T_L \to \infty}
 { \log {\cal A} \over - i T_L} =
 - \Xi(-2 i \mu) - \Xi(2 i\mu)
 }
We have interpreted the cutoff $L$ as the length of Lorentzian
time, $T_L = L$ since, for large $L$, this is the time it takes
for the {\it in} state to come out from the scattering region. In
other words the spatial cutoff $L$ is a good approximation for
times which are of order $L$.

 {}From the asymptotic expansion \asympU\ we can find the asymptotic
expansion of $\Xi$.
 \eqn\asyxi{
 \Xi(y)  = { 1 \over 2 \pi } \left[ ( \log {y \over 2} - {3 \over 2}){
 y^2\over 8} - {1\over 24} \log {y \over 2}
 -{7 \over 1440 y^2 } + \CO({1\over y^4}) \right]
 }
This implies that the only perturbative terms in the imaginary
part arise from the logarithms in \asyxi
 \eqn\imxi{ {\rm Im}[\Xi( - 2 i \mu)]  \approx   {|\mu| \mu \over 8}
 \left(1+{1\over 12}{1 \over \mu^2} \right)}
(Because of the dependence on the absolute value of $\mu$ they are
not analytic and should be kept.) Thus, going back to \obtainan ,
we find that the leading contribution to the real  part of the log
of the amplitude is
 \eqn\leadingim{\eqalign{
 {\rm Re}[ \log {\cal A} ] \approx & T_L \left[ -{ 1 \over 8}  |\mu
 + |\nu||(\mu+ |\nu|)\left(1+ {1\over 12}{1 \over
 (\mu+|\nu|)^2} \right) \right.\cr
 &\qquad \left.+ { 1 \over 8}   |\mu - |\nu||(\mu
 -|\nu|)\left(1+ {1\over 12}{1 \over (\mu-|\nu|)^2} \right) + \half
 \mu |\nu| \right] \cr
 = &\cases{ 0 & $\mu \pm |\nu| >0$ \cr
  T_L \, \mu |\nu| & $ \mu \pm |\nu| <0$}
 }}
So we see that for $\mu \pm | \nu| >0$ the leading approximation
to $\log {\cal A}$ is purely imaginary. Here we are in a
configuration where the two Fermi seas are above the barrier (see
figure 2 (c)/(d)), and the amplitudes is dominated by the leading
order transmission over the barrier, as expected. On the other
hand, for $\mu \pm |\nu| <0$, there is a negative real
contribution. This implies that this processes is suppressed. Here
the Fermi seas are below the barrier, and the in/out states are as
in figure 3 (a)/(b). (Figure 2(a)/(b) depict a configuration with
nonzero $\tilde \nu$.) In order to obey these boundary conditions
we need to have tunneling processes. We need of the order of
$|\nu|$ tunneling events per unit time. Each tunneling event
contributes a factor of $e^{ \mu \pi } $ (recall, $\mu <0$), which
is the contribution of a charged D-instanton ZZ brane. The number
of such factors depends on the total time $T_L$ as $ T_L |\nu|
/\pi$. This leads to \leadingim . Note that the effects of the
instantons do not exponentiate since we are looking at a very
special process where only a definite number of instantons could
contribute. Of course one could also study processes where
$\mu-|\nu|<0< \mu + |\nu|$.  Then, the second term in \imxi\ also
contributes.

The expression \obtainan\ for the partition function ${\cal A}$
has a few interesting consequences.  The two terms
$-\Xi(-2i(\mu +|\nu|) $ and $-\Xi(2i(\mu -|\nu|) $ can be
interpreted as the contribution of the fermions in the left and
the right side of the potential.  Therefore, either $-\Xi( 2i
\mu)$ or $-\Xi( -2i \mu)$ can be viewed as a nonperturbative
definition of the free energy of the bosonic $c=1$ system with
Fermi level $\mu$. (Recall, the bosonic system has fermions only
in one side of the potential.)  From this perspective the problem
with the $c=1$ system is that $-\Xi( \pm 2i \mu)$ are complex.
The sign ambiguity in the definition changes the sign of the
imaginary part which signals the instability of the system.

The expression \obtainan\ also gives an intuitive explanation of
the holomorphic factorization we have seen before.  Up to the
simple term which depends on $\mu|\nu|$ the partition function
factorizes as a product of $\exp(i L \Xi(-2i(\mu +|\nu|)$ and
$\exp(i L \Xi(2i(\mu -|\nu|)$ which are associated with the
incoming fermions from the left and the right side of the
potential. Our definition of the RR-flux is such that it does not
mix these two kids of fermions, and therefore we can specify
independent Fermi levels for them, $\mu \pm |\nu|$.  In Euclidean
space $|\nu| \to i |\nu|$ and therefore this separation explains
the holomorphic factorization we discussed above.  We will soon
add nonzero $\tilde \nu$, and will study the problem with a
Euclidean time circle.  The separation of these modes will
persist.  It underlies the holomorphic factorization of the
partition function.

Repeating this analysis for $\nu_{in} = -\nu_{out} = \tilde \nu$
we find an answer that is very similar to \obtainan\ except that
the last term changes sign
 \eqn\andnw{
 \log{\cal A}= -i  L  \left[-\Xi ( -2i(\mu + |\tilde \nu|)) - \Xi
 (2i(\mu - |\tilde \nu|)) -i  \half \mu  |\tilde\nu|  \right]
 }
In this case the real part is small for sufficiently large
negative $\mu$ but it is behaves as   $-  L  \mu  |\tilde\nu| $
for large positive $\mu$.  This is consistent with the duality
symmetry $\mu \to -\mu$, $\nu \leftrightarrow \tilde \nu$.

In the case that $\nu^2_{in} \not = \nu_{out}^2$ we have to insert
extra asymptotic states in order to balance the energy flux. We
will do this in more detail in the Euclidean computation in the
next subsection.

\subsec{Computation at finite $R$}

In this section we consider the finite temperature partition
function, where the time direction has period $\beta = 2 \pi R$.
Configurations with RR fluxes correspond to configurations where
the field $C$ or its dual have winding along the Euclidean time
direction. We have said that the field $C$ is at the self dual
radius, $C \sim C + 2 \pi \sqrt{2}$ with the normalization
\Ckinetic . This implies the following quantization condition for
the constant part of the Euclidean time derivative
 \eqn\implicq{
 \partial_\tau C = -i 2  \sqrt{2} \tilde \nu=  \sqrt{2} { \tilde q \over R}
 ~,~~~~~~~~~~~~\tilde q \in {\bf Z}
} We can similarly think about the quantization condition for,
$\tilde C$, the dual of $C$. This gives
 \eqn\implcdu{
 \partial_\phi C =  2 \sqrt{2} \nu = i \sqrt{2} {   q \over R}   ~,~~~~~~
  q \in {\bf Z}
 }
We then define
 \eqn\quantq{ q_{in} = { q +  \tilde q}
 ~,~~~~~~~~~~q_{out} = q- \tilde q
 }
Note that $q,\tilde q, q_{in}, q_{out}$ are integer, but $q_{in} -
q_{out}$ is always even. In these conventions an {\it in} right
moving fermion has $q_{in} = 1$. Note that we can consider {\it
in} states where we have a left moving spin field and a right
moving spin field. If the charge of the spin fields are opposite
this configuration gives us $q_{in}=1$. But then we should also
have  spin fields in the {\it out} state since \quantq\ \implicq\
\implcdu\ imply that $q^{in,out}$ are either both odd or both
even.

There are two closely related ways of thinking about the Euclidean
computation. One is to view it as an analytic continuation of the
Lorentzian scattering computation \amplic . The only difference is
that  the asymptotic regions now look like a cylinder. So we think
of the {\it in} and {\it out} states as living on a cylinder and
we expand them in fourier modes along the compact direction. The
analytic continuation of \mapinout\ gives the relation between
{\it in} and {\it out} fields. The resulting relation can be
summarized as (we continue to set $\alpha'=\half$)
 \eqn\condope{\eqalign{
 &\langle \psi^{out\dagger~ a }_{r} \psi^{in}_{b,~ - s} \rangle
 = \delta_{r , s} { \Gamma(\half+i \mu + s/R)\over\sqrt{2 \pi}}
 e^{{\pi\over 2}( \mu-i{ s\over R}) a b - i {\pi\over 4} ab} \cr
 &\langle \psi^{out}_{a,~r} \psi^{in\dagger~ b }_{-s} \rangle
 = \delta_{r,s}{ \Gamma(\half-i   \mu +s/R)\over\sqrt{2 \pi}}
  e^{{\pi\over 2}( \mu + i{ s\over R}) a b+ i {\pi\over 4} ab}
 }}
where $r,s \in {\bf Z} + \half $, $r,s >0$. When we do this
analytic continuation of \mapinout\ we might be a bit unsure about
the sign for $s$ in the right hand side. This sign is determined
by doing the analytic continuation of the fields carefully and
demanding that the mode, $\psi^{in}_{-s}$ does not annihilate the
vacuum, $\psi_{-s} |0 \rangle \not = 0$  for positive $s$ (in our
conventions $\psi^{in/out}_{s} |0\rangle =0 $ for $s >0 $). In
this description we are interested in computing an inner product
of the form
 \eqn\inppro{
 \langle \Psi_{out} | \Psi_{in} \rangle
 }
Notice that from the target space viewpoint, the states
$\Psi_{in}$ and $\Psi_{out}$ are determined by choosing the
non-normalizable behavior of the anti-holomporphic and the
holomorphic parts of the target space fields $T$ and $C$ near the
boundary. As usual, this correspondence involves bosonization of
the fermions and an identification of the modes of the bosonic
field with the modes of the $T$ and $C$ fields.

The other way of thinking about the problem consists in viewing
the problem as defined on a half cylinder where the {\it in} and
{\it out} fields are antiholomorphic and holomorphic fields
respectively.  We impose boundary condition at the asymptotic end
of the cylinder by specifying a state in the Hilbert space for
fermions on a cylinder. In the capped end of the semi infinite
cylinder we insert a boundary state which encodes the effects of
the scattering amplitude. This boundary state is also computed by
analytically continuing \mapinout . It is clear that both pictures
are equivalent and which one we choose is a matter of taste.

Let us consider a configuration with general $q_{in}$ and
$q_{out}$, and first consider the case
 \eqn\qasum{q_{in}\ge q_{out}\ge 0}
For simplicity, let us limit ourselves to $q_{in},q_{out} \in 2
{\bf Z}$. \foot{ This case is simpler because we do not need to
introduce spin fields.} So we will have a state in the {\it in}
Hilbert space of dimension $\Delta_{in}$. Since our problem is
invariant under translations in Euclidean time, we need that
$\Delta_{out} = \Delta_{in}$. We will be interested in considering
the state with lowest $\Delta_{in}$ since this is the state that
corresponds to exciting only the constant part of the RR field
strength (the gradient of $ (\partial_\phi - i \partial_t) C \sim
q_{in} $). This lowest dimension state is
 \eqn\dimin{ \Delta_{in}= {q_{in}^2 \over 4} = \Delta_{out} }
which corresponds to the state
 \eqn\psiqmin{ | \Psi\rangle_{in} =
 \prod_{l=1}^{q_{in}/2 } \psi^{ in \dagger +}_{ - l + 1/2}
   \psi_{ - , -l + \half }^{in} |0 \rangle
 }
States with higher dimension, with the same $q_{in}$ correspond to
exciting other oscillator modes of $T$ and $C$.

In the out Hilbert space we need a state with the same conformal
dimension. Since the S-matrix is given by the product of one body
S-matrices it is clear that we need as many fermions and holes in
the {\it out} Hilbert space as we have in the {\it in} Hilbert
space. Since our system is time translation invariant, the
amplitudes are diagonal in the mode number. So the non-zero
amplitudes have the form
 \eqn\ampio{\langle 0|  \prod_{l=1}^{q_{in}/2} \psi^{out\dagger~ a_n}_{l -\half}
 \psi^{out }_{b_n, l - \half}  \psi^{in\dagger +  }_{-l +\half}
 \psi^{in }_{- , -l + \half} | 0 \rangle
 }
Since the charge of the out state has to be $q_{out}$ we need that
$\sum b_n- a_n=  q_{out}$\foot{Notice that we are defining the
charge as the charge of the ket, which is minus the charge of the
{\it out} operators explicitly appearing in \ampio.}. There are
several ways to assign values of $a_n$ and $b_n$.    Of the
$q_{in}$ out-fermions, $q=(q_{in}+q_{out})/2$ should have ${out}$
charge minus one and $\tilde q=(q_{in}-q_{out})/2$ should have
${out}$ charge plus one. The number of possibilities of achieving
this is
 \eqn\nump{N(q,\tilde q)={q_{in}! \over
 [\half(q_{in}+q_{out})]![\half(q_{in}-q_{out})]!} ={(q+\tilde q)!
 \over q! \tilde q!}}
Using \condope\ we can compute \ampio\ and obtain

 \eqn\ampc{\eqalign{
 {\cal A}(\mu,q_{in},q_{out})
 & = e^{i \varphi(a_n,b_n, R)}
 \prod_{n=1}^{q_{in} /2}{|\Gamma(\half-i\mu +
 (n-\half)/R)|^2\over 2 \pi}e^{-\pi \mu (a_n- b_n)\over 2}
 {\cal A}(\mu,0,0)\cr
 & = e^{i \varphi(a_n,b_n, R)} e^{\pi \mu q_{out}\over 2}
 \prod_{n=1}^{q_{in}/2}{|\Gamma(\half-i\mu +
 (n-\half)/R)|^2\over 2 \pi} {\cal A}(\mu,0,0)\cr
 }}
where we have used  $q_{out}=\sum (b_n-a_n)$.  Note that up to the
phase $e^{i \varphi(a_n,b_n, R)}$ the answer does not depend on
the particular operator among all the $N(q_{in},q_{out})$
operators in \nump\ with the same charges\foot{ The technical
reason for this is the fact that the difference between the right
to right vs right to left amplitudes \mapinout\ is a simple
exponential.}.  It will be important below that this phase is
independent of $\mu$.

It is straightforward to extend the computation \ampc\ to values
of $q_{in}$ and $q_{out}$ which do not satisfy \qasum.  The answer
is expressed most easily in terms of $q$ and $\tilde q$
  \eqn\ampcg{
 {\cal A}(\mu,q_{in},q_{out}) =e^{i \varphi(a_n,b_n, R)}
  e^{ {\pi \mu \over 2} (|q|-|\tilde q|)}
  \prod_{n=1}^{\hat q/2 }{|\Gamma(\half+i\mu +
 {n-\half \over R})|^2\over 2 \pi} {\cal Z}_{0B}(\mu,q=\tilde q=0,R)
 }
where $\hatq = |q| + |\tilde q| =  q_{in}$ for the case \qasum .

Using \formla\ and the expression for $\Omega(y,r)$ \ndwdfr\ we
find for even $\hatq=2k$  \foot{ This is the same as the recursion
relations found for the function $G(x)$ in \fzz .}
 \eqn\OmegaC{\eqalign{
 \Omega(y={2k\over R} +2 i \mu,R) -& \Omega(y=0 +2i \mu,R) =\cr
 &=- \int_0^\infty { dt \over t}\left[ e^{ - i\mu t
 } { e^{ - {k t \over R}}-1 \over  4  \sinh   { t \over 2} \sinh
 { t\over 2R}}   +  {  k \over  t}  - ( {k^2\over2 R} +  i k \mu)
 e^{-t} \right]\cr
 &=\sum_{n=1}^k \int_0^\infty { dt \over t}\left[ { e^{ -
 ({2n-1\over 2R} + i \mu )t}  \over  2 \sinh
 { t\over 2}}   -  { 1 \over  t}   +  ( {2n-1 \over 2R}+ i \mu)
 e^{-t} \right]\cr
 &= \sum_{n=1}^k \log\left({\Gamma(\half+i\mu +
 {n-\half \over R} )\over \sqrt{2\pi}} \right)
 }}
Using this relation and the expression for ${\cal
Z}_{0B}(\mu,q=\tilde q=0)$ (see Appendix A), \ampcg\ can be
written as
 \eqn\zeoban{\eqalign{
 \log {\cal Z}_{0B}(\mu,q, \tilde q,R)=& \log {\cal
 A}(\mu,q_{in},q_{out})  = i \varphi(a_n,b_n, R)
  +{\pi \mu \over 2} (|q|-|\tilde q|) +\cr
  &+ \Omega(y={\hatq\over R} +2 i \mu,R) + \Omega(y={\hatq\over R}
  -2 i \mu,R)
 }}
which is our final expression for the free energy.

After we apply  T-duality
 \eqn\tduality{ R_B =  { \alpha' \over R_A} ~,~~~~~~~~~~~~~~~~ \mu_B =
 {R_A \over \sqrt{ 2 \alpha'} }\mu_A }
we find that \zeoban\ becomes the same as \result, up to the phase
and analytic terms in $\mu$ proportional to $\log R$. These terms
are related to the fact that we need to change the UV cutoff
$\Lambda$ when we perform T-duality (see the appendix).

Note that this 0B computation produces naturally the term that
involves $(|q| - |\tilde q|)$ while in the 0A problem we had to
introduce this term ``by hand" in order to match the expected
asymptotic behavior.

Notice that this procedure produces answers which are consistent
with T-duality, while previous studies did not.

The answer \ampcg\ has the expected symmetries: $A(\mu,q,\tilde
q)=A(\mu,-q,-\tilde q) = A(-\mu,\tilde q, q)=A(\mu,-q,\tilde
q)^*$. They follow from the two ${\bf Z}_2$ symmetries and time
reversal.

\bigskip
 \centerline{\bf Acknowledgements}
It is a pleasure to thank I.~Klebanov and E.~Witten for
discussions.  This work was supported in part by grant
\#DE-FG02-90ER40542.

\appendix{A} {Chiral Quantization of Matrix Models}

The purpose of this appendix is to derive some known results about
the $\hat c=1$ 0A and 0B matrix models.  We will use a formalism
which was first introduced in \AlexandrovFH\ and was later used
and elaborated on in
\refs{\KostovTK\AlexandrovPZ\AlexandrovQK\AganagicQJ%
\YinIV\AlexandrovKS\AlexandrovCG\TeschnerRD-\GaiottoGD}.  It
highlights the chiral nature of the problem and the scattering
from and to null infinities.  One of the advantages of this
formalism is that the theory is expressed in terms of {\it free
relativistic fermions}. The nontrivial scattering appears as a
nonlocal transform between the incoming and the outgoing
descriptions. The parabolic cylinder functions of the inverted
harmonic oscillator are replaced by simple wave functions in the
$p\pm x$ representation. $p\pm x$ are the analog of creation and
annihilation operators of the ordinary harmonic oscillator and
their eigenstates are analogous to the familiar coherent states.
However, unlike the ordinary harmonic oscillator, since $p\pm x$
are {\it hermitian} operators, their eigenvalues are real and the
inner product of functions in these representations is standard.

We will present this formalism, will clarify some of its
properties and will extend it.  We will start the discussion of
the first quantized theories with some general properties of
eigenstates of $p\pm x$, and will then use them in the special
cases relevant to the 0B and 0A strings. Then, we will study the
second quantized theories and will compute their free energies.

\subsec{First quantized problems}

\bigskip\centerline{\it A single upside down harmonic oscillator}

Consider first a generic quantum mechanical problem of a single
degree of freedom. Standard bases of orthonormal states are
$|x\rangle$ and $|p\rangle$, which are coordinate and momentum
eigenstates respectively, with $\langle x |p \rangle = {1 \over
\sqrt{2 \pi}} e^{ipx}$.  We will also be interested in the bases
$|s\rangle$ and $|u\rangle$ which are orthonormal eigenstates of
 \eqn\SUdef{\eqalign{
 &S ={ P + X \over \sqrt 2} \cr
 &U = {P - X \over \sqrt 2} \cr
 & [S,U]=i}}

\medskip \ifig\figphase{ The phase space for a harmonic oscillator
parameterized by the $x,p$ coordinates or the $s,u$ coordinates.
The dotted lines denote various possible classical trajectories.}
{\epsfxsize=0.4\hsize\epsfbox{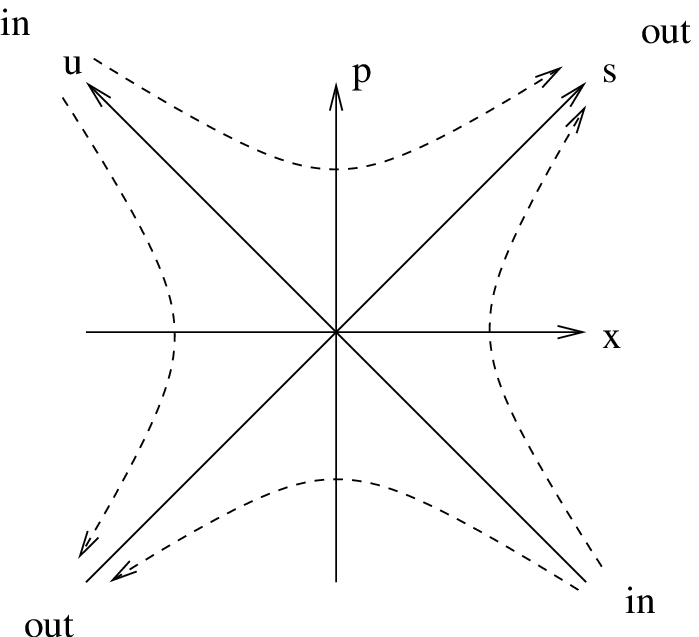}}

 It is easy to find the inner products
 \eqn\uswave{\eqalign{
 &\langle x|s\rangle ={2^{1\over 4} e^{ i {\pi \over 8} }\over \sqrt{2\pi}}
 \exp\left(i(-{x^2 \over 2} +\sqrt 2 sx - {s^2\over 2})\right) \cr
 &\langle x|u\rangle= {2^{1\over 4} e^{ - i {\pi \over 8} }  \over \sqrt{2\pi}}
 \exp\left(i({x^2\over 2} +\sqrt 2 ux+  {u^2\over 2})\right) \cr
 & \langle s|u\rangle ={ 1  \over \sqrt{ 2\pi} } \exp\left(isu\right)
 }}
The $s$ and $u$ dependent phases in $ |s \rangle $ and $|u\rangle$
are such that $U=S- \sqrt 2 X $ acts on $ \langle s|u\rangle $,
and $\langle s|x \rangle $ as $-i \partial_s$ and similarly for
the action of $S$  on $ \langle u|s\rangle $ and $ \langle u|x
\rangle$.  In the last expression we defined the integral $\int dx
e^{i x^2}$ as  $\int dx e^{(i-0^+) x^2}=\sqrt{i \pi}$. We have
chosen the constant phases of the first two lines so as to
simplify the last line and some of the subsequent formulas.

So, let us now focus on the inverted harmonic oscillator with the
Lagrangian and Hamiltonian
 \eqn\flag{\eqalign{
 &L= \half(\dot X^2 + X^2)\cr
 &H= \half(P^2 - X^2)= {1\over 2}(S  U + US) }}
It is easy to work out the time evolution
 \eqn\timeev{\eqalign{
 &e^{-iHt} |s\rangle = e^{{t\over 2}} |e^{t} s \rangle \cr
 & e^{-iHt} |u\rangle = e^{-{t\over 2}} |e^{-t} u \rangle \cr
  &\langle s| e^{-iHt}  = e^{-{t\over 2}} \langle e^{-t} s | \cr
  &\langle u| e^{-iHt}  = e^{{t\over 2}} \langle e^{t} u | \cr
 & \langle s| e^{-iHt}|u\rangle = { 1 \over \sqrt{  2 \pi} }
  e^{-{t\over 2}}\exp\left(isue^{-t}\right)}}
Here the factors of $e^{\pm {t\over 2}}$ are needed for unitarity,
but also come out of \flag\ by writing, in the $s$ basis $H = - i
s \partial_s - { i \over 2} $, and $H= i u \partial_u  +{ i \over
2}$ in the $u$ basis.

The operators \SUdef\ are similar to the creation and annihilation
operators of the ordinary (or ``upside up") harmonic oscillator.
One difference is that in our case, $S$ and $U$ are hermitian
operators and not hermitian conjugates to each other.
Correspondingly the states $|s\rangle$ or $|u\rangle$ are
analogous to coherent states. Since these operators are hermitian
the states $\langle s|$ and $\langle u |$ will also eigenstates of
$S$ and $U$ respectively. These two bases will be useful to
describe the initial and final states of the upside down harmonic
oscillator. In other words, the incoming states will be naturally
described in terms of the $u$ basis and the outgoing states in
terms of the $s$ basis. This can be seen quite naturally by
looking at the shape of trajectories in \figphase , but will be
seen more precisely later.

There are two linearly independent energy eigenstates for every
energy $\epsilon$. In the $x$ representation the wavefunctions are
the two parabolic cylinder functions. They can be taken to be even
and odd under the parity transformation $x \to -x$. Alternatively,
we can take one wavefunction to correspond to a wave coming from
the left and scattered to the right and back to the left, and the
other wave function obtained from this one by $x \to -x$.

In the $s$ representation the energy eigenstates with eigenvalue
$\epsilon$ are $s^{i\epsilon-\half}$.
 The
singularity at $s=0$ leads to a two fold doubling of the number of
states $|\epsilon,out \pm \rangle$, where the label $out$ will be
explained shortly.   Their wavefunctions are
 \eqn\psispm{\eqalign{
 &\langle s|\epsilon,out + \rangle=
 \cases{{1 \over \sqrt{2\pi}}s^{i\epsilon-\half}& $s>0$\cr 0 & $s<0$}\cr
 &\langle s|\epsilon,out - \rangle=
 \cases{0& $s>0$\cr {1 \over \sqrt{2\pi}}(-s)^{i\epsilon-\half} & $s<0$}
 }}
By looking at the trajectories in \figphase\ we see that
$|\epsilon, out +\rangle $ states are states that in their
outgoing modes contain only a right moving piece. While the states
$|\epsilon, out - \rangle$ contain only a left moving piece in
their outgoing modes. Therefore we will refer to them as ``{\it
out} states".

Another natural basis  arises from the $u$ representation
 \eqn\psiupm{\eqalign{
 &\langle u|\epsilon,in + \rangle=
 \cases{{1 \over \sqrt{2\pi}}u^{-i\epsilon-\half}& $u>0$\cr 0 & $u<0$}\cr
 &\langle u|\epsilon,in - \rangle=
 \cases{0& $u>0$\cr {1 \over \sqrt{2\pi}}(-u)^{-i\epsilon-\half} & $u<0$}
 }}
These are states which contain only right/left moving incoming
pieces for $+/-$. We will refer to them as ``{\it in} states".

Semiclassically the incoming states have $u \to \pm \infty$ and $s
\approx 0$, while the outgoing states have $u \approx 0$ and $s\to
\pm \infty$.  Therefore, it is natural to take the incoming states
to be $|\EE,in \pm \rangle$, where $|\EE,in + \rangle$ describes a
particle coming from the left (negative $x$) and $|\EE,in -
\rangle$ describes a particle coming from the right (positive
$x$). Similarly, the outgoing states are $| \EE,out \pm \rangle$.
Here, $|\EE,out + \rangle$ describes a particle going to the right
(positive $x$), and $|\EE,out - \rangle$ describes a particle
going to the left (negative $x$).

 These two bases are related by a unitary transformation
 \eqn\chanb{\eqalign{
 &\pmatrix {| \epsilon,out + \rangle \cr | \epsilon,out -\rangle}
 ={\cal S} \pmatrix {| \epsilon,in + \rangle \cr | \epsilon,in -\rangle} \cr
 &{\cal S} =\pmatrix{{ e^{ i {\pi \over 4}} e^{\pi \epsilon \over 2}
 \over \sqrt{2 \pi}}\Gamma(\half-i \epsilon ) & { e^{ -i {\pi \over 4}}e^{-{\pi \epsilon \over 2}}
 \over\sqrt{2 \pi}}\Gamma(\half-i \epsilon ) \cr
 { e^{ - i {\pi \over 4}}e^{-{\pi \epsilon \over 2}} \over\sqrt{2 \pi} }\Gamma(\half-i \epsilon )&
 {e^{ i {\pi \over 4}} e^{\pi \epsilon \over 2} \over \sqrt{2 \pi}}\Gamma(\half-i\epsilon )}=
 \pmatrix{{ e^{ i {\pi \over 4}} e^{i\Phi_B(\epsilon )}\over \sqrt{1 + e^{-2\pi \epsilon }}} &
 {e^{ - i {\pi \over 4}} e^{i\Phi_B(\epsilon )}\over \sqrt{1 + e^{2\pi \epsilon }}}\cr
 {e^{ -i {\pi \over 4}}e^{i\Phi_B(\epsilon )}\over \sqrt{1 + e^{2\pi \epsilon }}}&
 { e^{ i {\pi \over 4}} e^{i\Phi_B(\epsilon )}\over \sqrt{1 + e^{-2\pi \epsilon }}}} \cr
 &e^{i\Phi_B(\epsilon )}=  \sqrt{\Gamma(\half-i\epsilon)\over
 \Gamma(\half+i\epsilon )}
 }}
Here we wrote $\langle \epsilon, out + | \EE, in- \rangle = \int
ds du \langle \EE , out + |s\rangle \langle s | u \rangle \langle
u | \EE, in - \rangle$ and we used \psispm , \psiupm , \uswave .
   Another way to understand these bases
is to express the $out$ states as functions of $u$ and the $in$
states as functions of $s$.  Using \chanb, or more directly by
using $\langle s | u \rangle$ in \uswave\ and Fourier transforming
\psispm\psiupm, we find
 \eqn\suwavef{\eqalign{
 &\langle s |\EE,in + \rangle =
 {e^{ i {\pi \over 4}} e^{i\Phi_B(\EE)}\over \sqrt{2\pi} \sqrt{1 + e^{-2\pi \EE}}}
  (s+i 0^+)^{i\EE-\half}\cr
 &\langle s |\EE,in - \rangle =
 { e^{- i {\pi \over 4}}e^{i\Phi_B(\EE)}\over \sqrt{2\pi} \sqrt{1 + e^{2\pi \EE}}}
  (s+i 0^-)^{i\EE-\half}\cr
 &\langle u |\EE,out + \rangle =
 { e^{ -i {\pi \over 4}} e^{-i\Phi_B(\EE)}\over \sqrt{2\pi} \sqrt{1 + e^{-2\pi \EE}}}
  (u + i 0^-)^{-i\EE-\half}\cr
 &\langle u |\EE,out - \rangle =
 { e^{ i {\pi \over 4}} e^{-i\Phi_B(\EE)}\over \sqrt{2\pi} \sqrt{1 + e^{2\pi \EE}}}
  (u+i0^+)^{-i\EE-\half}\cr
 }}
where the $s +  i 0^\pm  $ prescription means that for negative
$s$ we substitute $s^{i \alpha} = ( |s| e^{\pm  i \pi} )^{i
\alpha} = (-s)^{i \alpha} e^{ \mp \pi \alpha}$. So the $|\EE ,
in\pm \rangle $ states in the $s$ representation $\langle s |\EE ,
in\pm \rangle $ are linear combinations of  $\langle s|\EE, out
\pm \rangle $ whose precise coefficients are determined by
thinking of $\langle s|\EE , in \pm \rangle$ as functions that are
analytic in the upper/lower half $s$ plane. The situation is very
similar to one that arises in the physics of Rindler space when we
express the Minkowski wavefunctions in terms of the Rindler
wavefunctions. In our case this arises from the fact that
$|\epsilon , in + \rangle $ has support only for $u>0$, this
implies that in the $s$ representation this is an analytic
function for $Im(s)>0$. In the Rindler case, the positive
frequency condition on the Minkowski wavefunctions implies similar
analyticity properties. In  \bruno\ a connection between these
fermions and  fermions in de-Sitter space was studied. In
\FriessTQ\ the thermal looking nature of these amplitudes was
explored.

The even and odd states $|\EE,out + \rangle \pm |\EE,out - \rangle
$ and $|\EE,in  + \rangle \pm |\EE,in - \rangle $ diagonalize
\chanb
 \eqn\diagA{\eqalign{
 &| \EE,in  + \rangle \pm | \EE,in -\rangle = e^{i \varphi_\pm (\EE)}
 \left( | \EE,out + \rangle \pm | \EE,out -\rangle \right)\cr
 &e^{i \varphi_+ (\EE)}= e^{i\Phi_B(\EE)}{ e^{  i { \pi \over 4} } +
  e^{ - i { \pi \over 4} }
 e^{-\pi \EE} \over
 \sqrt{1+e^{-2 \pi \EE}}}= 2^{-i \EE}  {\Gamma({1\over 4}
 - {i \EE \over 2})\over \Gamma({1\over 4 } + {i \EE \over 2})}\cr
  &e^{i \varphi_- (\EE)}=
  e^{i\Phi_B(\EE)}{{ e^{  i { \pi \over 4} } - e^{ - i { \pi \over 4} }
   e^{-\pi \EE}  \over
 \sqrt{1+e^{-2 \pi \EE}}}= 2^{-i \EE} i {\Gamma({3\over 4}
 - {i \EE \over 2})\over \Gamma({3\over 4 } + {i \EE \over 2})}}}}

  Our interpretation of \psiupm \psispm\ as {\it in } and {\it out}
  states   is further supported
by comparing \psispm\psiupm\ and \suwavef.  The {\it in}
 states $|\EE ,in
\pm \rangle$ have support only for one sign of $u$ and for both
signs of $s$.  This is the expected behavior of incoming states.
The $out$ states $|\EE,out \pm \rangle$ have support only for one
sign of $s$ and for both signs of $u$.  This is the expected
behavior of outgoing states. Furthermore, for large $|\EE|$ the
relation between the two bases is simple.  Up to an $\EE$
dependent phase $|\EE,in \pm \rangle \sim |\EE,out \pm \rangle$
for positive $\EE$ and $|\EE,in \pm \rangle \sim |\EE,out \mp
\rangle$ for negative $\EE$.  This is consistent with the
semiclassical picture of complete transmission for positive $\EE$
and complete reflection for negative $\EE$.

One can actually  show more precisely why it is reasonable
associate the basis $|\EE, in \pm \rangle$ with incoming states.
For that purpose we can compute $\langle x | \EE, in \pm \rangle $
using \uswave\ \psiupm\ . We do not need the exact answer, which
is a combination of parabolic cylinder functions. We only need the
behavior of the function for large $x$. Since $x$ is large we can
compute the answer by saddle point integration. The saddle point
equation for $u$ is $ \sqrt{2} x + u - \EE/u \sim 0$. The two
saddle points, at $u \sim  -\sqrt{2 } x$ and at $ u \sim
\EE/(\sqrt{2} x)$,  give the incoming and outgoing pieces of the
$x$ space wavefunction, which go like  $ e^{ - i {x^2 \over 2} } $
and $e^{ i { x^2 \over 2} }$ to leading order in $x$. Note that
the first saddle point arises only for $ \pm x <0$ for $|\EE, in
{\pm }\rangle$. This means that the incoming wavefunction is
supported to the left/right side of the potential for $|\EE, in
{\pm} \rangle$. Furthermore, the coefficient of the first saddle
point is energy independent (except for a simple, expected, factor
of $|x|^{- i \EE -\half}$). This is the natural normalization for
the incoming states. The integral in the region close to the
second saddle point gives us the reflected part of the
wavefunction and contains the information about scattering phase.

Repeating this discussion for $| \EE, out \pm \rangle $ we can
understand why it is natural to associate them to outgoing states
which are right or left moving.

\eject

\bigskip\centerline{\it Two upside down harmonic oscillators}

\bigskip

Now, let us discuss the same problem but with two degrees of
freedom $X_1$ and $X_2$ and their conjugate momenta $P_1$ and
$P_2$.  We change variables to polar coordinates $X_1=X \cos
\theta$, $X_2 = X \sin\theta$.  The momentum conjugate to $\theta$
is
 \eqn\qdef{q=X_1P_2 - X_2 P_1}
and we can work in a sector where it is a fixed integer $c$
number.  Then, we have two natural bases of states $|x \rangle$
and $|p \rangle$ which are eigenstates of $X$ and
 $P=\cos\theta P_1 + \sin\theta P_2$ respectively. Note that, even though
 $(X_i ,P_i)$ are canonically conjugate, $P$ is not the momentum conjugate
 to $X$.

As in \SUdef, we define
 \eqn\SUdef{\eqalign{
 &S_i ={ P_i + X_i \over \sqrt 2} \cr
 &U_i = {P_i - X_i \over \sqrt 2} \cr
 & [S_i,U_j]=i\delta_{i,j}}}
and we can again change to ``polar coordinates''
\eqn\SUdef{\eqalign{
 &S_1 =S \cos \theta_s \cr
 &S_2 =S \sin \theta_s \cr
 &U_1 =U \cos \theta_u \cr
 &U_2 =U \sin \theta_u \cr
 }}
The momenta conjugate to $S$ and $U$ are $P_s$ and $P_u$.  It is
important that they are not given by $U$ and $-S$.  However, $q$
of \qdef\ can be written also as $q= S_1U_2-S_2U_1$, and it is the
momentum conjugate to both $\theta$, $\theta_s$ and $\theta_u$.
This follows from the fact that this is the charge of the same
rotation symmetry.

Let us study the various bases in more detail.  The simplest
states are $X_i$ eigenstates, $|x_1,x_2\rangle$, or in polar
coordinates $|x,\theta\rangle$. Note that the eigenvalue $x$ is
positive. They satisfy
 \eqn\xieig{\langle x_1,x_2|x,\theta\rangle = \sqrt x
 \delta(x_1 - x \cos \theta)\delta(x_2-x \sin\theta)}
Instead of diagonalizing $\theta$ it is better to diagonalize $q$,
$|x,q\rangle = {1\over \sqrt{2\pi}} \int d\theta e^{iq\theta}
|x,\theta\rangle$.  We will use similar notation for various bases
diagonalizing the $S$ or $U$ variables (again, the eigenvalues $s$
and $u$ are positive). One way to find the inner products between
these bases is to convert to the bases where the Cartesian
coordinates $X_i$, $S_i$ or $U_i$ are diagonal and then use
\uswave\ for each of them.  We readily find (recall,
$\int_0^{2\pi} d \theta e^{i q\theta} e^{i a \cos\theta}=2 \pi i^q
J_q(a)$)
 \eqn\suxqi{\eqalign{
 \langle s, \theta_s| x, q \rangle &={e^{- i {\pi \over 4}} i^{- 3 q \over 2}  \sqrt{2 x s} \over
 (2\pi)^{3\over 2}}\int_0^{2\pi} d \theta e^{iq\theta}
 \exp\left(i({x^2 \over 2} -\sqrt 2 sx \cos (\theta - \theta_s) +
 {s^2\over 2})\right)\cr
 & = { e^{- i {\pi \over 4}} (-1)^q i^{ - { q \over 2}} \sqrt{2 x s} \over \sqrt{2\pi}} \exp\left(i q
 \theta_s+ i{x^2\over 2}+i{s^2\over 2}\right) J_q(\sqrt 2 sx)
 \cr
 \langle u, \theta_u| x, q \rangle &={e^{ i {\pi \over 4}} i^{- { q \over 2} } \sqrt{2x u}\over
 (2\pi)^{3\over 2}}\int_0^{2\pi} d \theta e^{iq\theta}
 \exp\left(i(-{x^2 \over 2} -\sqrt 2 ux \cos (\theta - \theta_u) -
 {u^2\over 2})\right)
 \cr
 & =  {e^{ i {\pi \over 4}} (-1)^q i^{q \over 2} \sqrt{2 x u} \over \sqrt{2\pi}}
 \exp\left(i q \theta_u-i{x^2 \over 2} - i{u^2\over 2}
 \right)J_q(\sqrt 2 ux)\cr
 \langle s, q | u,q' \rangle &={i^{-q} \delta_{q,q'}\sqrt{su}\over 2\pi}
 \int_0^{2\pi} d\theta_u e^{i q \theta_u + i s u \cos \theta_u}
 =   \sqrt{su}J_q(su) \delta_{q,q'}
 }}
where   we have chosen the overall phases to simplify some
formulas later. The first two expressions demonstrate that $q$ is
the momentum conjugate to $\theta$, $\theta_s$ and $\theta_u$. The
last inner product can be derived in several different ways whose
consistency relies on (or better, gives a proof of) the Weber's
formula
 \eqn\weber{\int_0^\infty e^{-p x^2} J_q(ax)J_q(bx) xdx=
 {e^{-{ a^2+b^2 \over 4 p}} \over 2 p} I_q\left({ ab \over 2p}\right) ={e^{-{a^2+b^2 \over 4
 p}} \over 2 p} i^{-q}J_q\left({ i ab \over 2p}\right)}
with $p=-i+0^+$.

We now study the system with the Hamiltonian
 \eqn\twocH{\eqalign{
 H&=\half (P_1^2 + P_2^2 - X_1^2 - X_2^2)=
 \half (P^2 + {q^2 - {1 \over 4} \over X^2} - X^2) \cr
 &= \half (S_1U_1 + U_1 S_1 +  S_2U_2 + U_2 S_2)\cr
 &=\half( S P_s + P_s S)\cr
 &= - \half( U P_u + P_u U)
 }}
We will take $q$ to be a $c$ number and will view the system as
having a single degree of freedom. This Hamiltonian has two
natural energy eignestates $|\EE,in \rangle$ and $|\EE,out\rangle$
with wavefunctions and inner products
 \eqn\sutw{\eqalign{
 &\langle s|\EE ,out\rangle = {1 \over \sqrt {2 \pi}} s^{i\EE  -\half} \cr
 &\langle u|\EE ,in\rangle = {1 \over \sqrt {2 \pi}} u^{-i\EE  -\half}\cr
 &\langle \EE ,out|\EE ',in\rangle = e^{i \Phi_A(\EE )} \delta(\EE -\EE ')=
 2^{-i\EE } {\Gamma\left(\half(1 + q - i\EE )\right) \over
 \Gamma\left(\half(1 + q +i\EE )\right)}\delta(\EE -\EE ')\cr
 }}
where in the last inner product we used the integral
$\int_0^\infty dy y^{-i\EE } J_q(y) =2^{-i\EE }
{\Gamma\left(\half(1 + q - i\EE )\right) \over \Gamma\left(\half(1
+ q +i\EE )\right)}$ with nonnegative $q$.  Note, as a check that
these inner products are independent of the sign of $q$.

\subsec{Second quantized problem}

\bigskip\centerline{\it 0B}

The 0B matrix model is a system of fermions whose first quantized
description is the first problem discussed above.  Its Lagrangian
is
 \eqn\zbla{{\cal L}=\int_{-\infty}^{\infty} dx \Psi^\dagger(x,t)
 \left(i \partial_t +\half \partial_x^2 +\half x^2 + \mu\right)
 \Psi(x,t) }
In order to express it in terms of the $s$ and $u$ variables we
define the fermionic fields
 \eqn\psiusd{\eqalign{
 &\Psi_s(s,t)=\int dx  \langle s | x \rangle \Psi(x,t) = {2^{1\over 4}
 e^{ - i { \pi \over 8}} \over \sqrt{2\pi}
 }\int_{-\infty}^\infty dx \exp\left(i({x^2\over 2} -\sqrt 2 sx +
 {s^2\over 2})\right) \Psi(x,t) \cr
 &\Psi_u(u,t)= \int d x \langle u | x \rangle \Psi(x,t)=  {2^{1\over 4}
  e^{  i { \pi \over 8}}\over \sqrt{2\pi}
 }\int_{-\infty}^\infty dx \exp\left(i(-{x^2\over 2} - \sqrt 2 ux
 - {u^2\over 2})\right) \Psi(x,t)
 }}
  which are related through
a Fourier transform
 \eqn\psisur{\Psi_s(s,t)=\int d u  \langle s | u \rangle \Psi_u(u,t)
 = {1\over \sqrt{2\pi}}
 \int_{-\infty}^\infty du \exp\left(isu \right) \Psi_u(u,t)}
and find
 \eqn\zblasu{\eqalign{
 {\cal L}&=\int_{-\infty}^{\infty} ds \Psi_s^\dagger(s,t)\left(i
 \partial_t  +{i \over 2}(s \partial_s +\partial_s s) + \mu
 \right)  \Psi_s (s,t)\cr
 &=\int_{-\infty}^{\infty} du \Psi_u^\dagger(u,t)\left(i
 \partial_t -{i \over 2} (u \partial_u +\partial_u u) + \mu\right)
 \Psi_u(u,t)
 }}
We can further simplify the analysis by the change of variables
 \eqn\psir{\eqalign{
 &\Psi^{(in)}_1(r,t)=e^{r\over 2} \Psi_u(u=e^r,t) \cr
 &\Psi^{(in)}_2(r,t)=e^{r\over 2} \Psi_u(u=-e^r,t) \cr
 &\Psi^{(out)}_1(r,t)=e^{r\over 2} \Psi_s(s=e^r,t) \cr
 &\Psi^{(out)}_2(r,t)=e^{r\over 2} \Psi_s(s=-e^r,t)
 }}
which makes the Lagrangians \zblasu\ look relativistic
 \eqn\zblar{\eqalign{
 {\cal L}&=\int_{-\infty}^{\infty} dr \sum_{i=1,2}
 \Psi^{(in)\dagger}_i (r,t)\left(i
 \partial_t  -i  \partial_r  + \mu \right)  \Psi^{(in)}_i(r,t)\cr
 &=\int_{-\infty}^{\infty} dr \sum_{i=1,2}
 \Psi^{(out)\dagger}_i(r,t)\left(i
 \partial_t  +i  \partial_r  + \mu \right)  \Psi^{(out)}_i(r,t)\cr
 }}
The parameter $\mu$ is like the time component of a vector field
coupled to the fermion number current whose incoming and outgoing
components are
 \eqn\fermcurr{\eqalign{
 &J^{(in)}=\sum_i\Psi^{(in)\dagger}_i \Psi^{(in)}_i \cr
 &J^{(out)}=\sum_i\Psi^{(out)\dagger}_i \Psi^{(out)}_i
 }}
We can remove it by a time dependent gauge transformation, but we
prefer not to do so. In this form it is clear that we have four
incoming Majorana Weyl fermions $\Psi^{(in)}$ and four outgoing
Majorana Weyl fermions $\Psi^{(out)}$ of the opposite chirality.
The incoming and the outgoing fermions are related through our map
\psisur\ which becomes
 \eqn\mapinout{\eqalign{
 & \Psi^{(out)}_1(r,t)= { 1 \over \sqrt{2 \pi} } \int_{-\infty}^\infty dr'
 e^{\half(r+r')}\left(\exp(i e^{r+r'})\Psi^{(in)}_1(r',t)+
 \exp(-i e^{r+r'})\Psi^{(in)}_2(r',t) \right)\cr
 & \Psi^{(out)}_2(r,t)= { 1 \over \sqrt{2 \pi} } \int_{-\infty}^\infty dr'
 e^{\half(r+r')}\left(\exp(-i e^{r+r'})\Psi^{(in)}_1(r',t)+
 \exp(i e^{r+r'})\Psi^{(in)}_2(r',t)\right)
 }}

What are the symmetries of our problem?  In the form \zblar\ the
Lagrangian has an incoming $SO(4)\approx SU(2) \times SU(2)'$
symmetry which rotates the incoming fermions and similarly an
outgoing $SO(4)$ symmetry which rotates the outgoing fermions. The
coupling to $\mu$ breaks each of these symmetries to $SU(2)\times
U(1)$.  The map between the incoming and outgoing fields
\mapinout\ breaks most of these symmetries. But, let us first
ignore the map and start, without loss of generality, by
considering the incoming symmetry.  The current of the $U(1)$
factor has already been mentioned in \fermcurr.  The incoming
$SU(2) $ currents are
$J^{(in)+}=\Psi^{(in)\dagger}_1\Psi^{(in)}_2$,
$J^{(in)-}=\Psi^{(in)\dagger}_2\Psi^{(in)}_1$ and
$J^{(in)0}=\Psi^{(in)\dagger}_1\Psi^{(in)}_1
-\Psi^{(in)\dagger}_2\Psi^{(in)}_2$.  We note that the currents
$J^{(in)}$ of \fermcurr\ and $J^{(in)0}$ are local in our original
``space'' coordinate $s$, while the currents $J^{(in)\pm}$ are
nonlocal.  The latter involve creating a fermions at $s$ and
annihilating a fermion at $-s$, or the other way around.  The same
distinction between these currents applies in the $x$ coordinate.

These four currents have an obvious string theory interpretation.
After bosonization the fermion number current $J^{(in)}$ creates
an incoming NS-NS tachyon $T^{(in)}$; roughly\foot{We use the word
``roughly'' because of the non-local transform  between $\phi$ and
$r$.
 } $J^{(in)}\sim (\partial_t-\partial_\phi) T^{(in)}$. The
current $J^{(in)0}$ creates the incoming R-R scalar $C^{(in)}$;
roughly $J^{(in)0}\sim (\partial_t-\partial_\phi) C^{(in)}$. These
two excitations, which are local in $x$, correspond to the
perturbative string spectrum. The other two currents $J^{(in)\pm}$
create nonperturbative string states. These are solitons --
coherent states of an infinite number of $C^{(in)}$ quanta;
roughly $J^{(in)\pm}\sim e^{\pm i \sqrt 2 C^{(in)}}$. Such
solitons were studied in \DeWolfeQF . They create a fermion at one
sign of $x$ and annihilate it at the other. Note that this is
consistent with the $C$ field being at the $SU(2)$ radius.  We now
interpret this $SU(2)$ symmetry as rotating the two fermion
flavors in \zblar.

This discussion of the incoming symmetries is trivially repeated
for the outgoing symmetries. In terms of the field $C$ the $SU(2)$
symmetries of the past and the future are simply those of the left
moving and the right moving fields at the selfdual radius.

The map from the past to the future \mapinout\ shows that only one
of the two $U(1)$ symmetries is conserved -- the total incoming
fermion number equals the total outgoing fermion number. The two
$SU(2)$ symmetries are more interesting.  Both of them are broken,
but for large $|\mu|$ a certain $U(1) \subset SU(2)^{(in)} \times
SU(2)^{(out)}$ is approximately conserved. It is broken only by
nonperturbative effects of order $e^{-c |\mu|}$ for some constant
$c$. The physical interpretation of this fact is simple. We start
with a vacuum with $N\to \infty$ fermions. Let us prepare an
initial state with $\half N + n^{(in)}$ incoming fermions from
negative $x$ and $\half N - n^{(in)}$ incoming fermions from
positive $x$, and let us examine a final state with $\half N +
n^{(out)}$ outgoing fermions to positive $x$ and $\half N -
n^{(out)}$ outgoing fermions to negative $x$. For $\mu\to -
\infty$, where the fermions are far below the barrier, there is
almost no communication between the left and right sides of the
potential, and we must have $n^{(in)} \approx - n^{(out)}$.
Conversely, for $\mu \to + \infty$ we have $n^{(in)} \approx
n^{(out)}$. But for finite $\mu$ we can have arbitrary $n^{(in)}$
and $n^{(out)}$. Such scattering processes are created with the
insertion of $n^{(in)}$ insertions of $e^{i \sqrt 2 C^{(in)}}$ in
the past (for negative $n^{(in)}$ we take $e^{-i \sqrt 2
C^{(in)}}$), and $n^{(out)}$ insertions of $e^{i\sqrt 2
C^{(out)}}$ in the future. The condition $n^{(in)}\approx
n^{(out)}$ for $\mu\to +\infty$ states that the winding  of $C$
  is approximately conserved while the momentum of $C$
is not conserved. For $\mu \to - \infty$ we have the reverse
situation.

Let us discuss the discrete symmetries of our problem. First, a
${\bf Z}_2$ subgroup of the the $SU(2)$ we mentioned above is not
broken by the map \mapinout.  Combining it with a $U(1)$
transformation we can identify it as the parity transformation $x
\to -x$, $s\to -s$ and $u\to -u$.  In terms of the in and out
fermions its action is $\Psi^{(in/out)}_1(r,t) \leftrightarrow
\Psi^{(in/out)}_2(r,t)$.  We identify this transformation with the
spacetime charge  conjugation which is generated by the worldsheet
transformation $(-1)^{F_L}$ ($F_L$ is the leftmoving spacetime
fermion number).  As a check, note that the currents
$J^{(in/out)}$ are even and the currents $J^{(in/out)\pm}$ are odd
under this transformation. Note that spacetime charge conjugation
is parity in the matrix model.

Let us consider now the  S-duality symmetry in spacetime, which is
generated in the worldsheet description by $(-1)^{f_L}$ ($f_L$ is
the leftmoving spacetime fermion number).  It acts on the
parameter $\mu$ as $\mu \to -\mu$, and therefore it is a symmetry
only for $\mu=0$. Its action on the fields is
 \eqn\ztwosy{\eqalign{
 &\Psi(x,t) \to   {1 \over \sqrt{2\pi}}
 \int_{-\infty}^\infty dx' e^{ix'x} \Psi^\dagger(x',t)\cr
 &\Psi_s(s,t) \to  \Psi_s^\dagger(s,t)\cr
 &\Psi_u(u,t) \to  \Psi_u^\dagger(-u,t)\cr
 &\Psi^{(in)}_1 \leftrightarrow \Psi^{(in)\dagger}_2 \cr
 &\Psi^{(out)}_i \to \Psi^{(out)\dagger}_i
 }}
It is easy to check that it is a symmetry of the Lagrangian
\zbla\zblasu\zblar\ with $\mu=0$, and of the transforms
\psiusd\psisur\mapinout. Clearly, if we combine this operation
with the parity transformation $(-1)^{F_L}$ the difference between
the transformations of $\Psi_s$ and $\Psi_u$ and the difference
between $\Psi^{(in)}$ and $\Psi^{(out)}$ are reversed. It is
interesting that while in $x$ space the transformation involves a
duality transformation of $(x,p)$, which is implemented by a
Fourier transform, in $s$, $u$ and $r$ space no such duality
transformation is needed, and the transformation rules are local
and simple. We conclude that in the $s$, $u$ and $r$ variables
this ${\bf Z}_2$ symmetry acts as charge conjugation on the
outgoing fields and as $CP$ on the ingoing fields.

We point out that this ${\bf Z}_2$ symmetry is a subgroup of the
original $SO(4)$ we mentioned above.

\bigskip\centerline{\it 0A}

The 0A matrix model is a system of fermions whose first quantized
description is the second problem discussed above.  Its Lagrangian
is
 \eqn\zala{{\cal L}=\int_{0}^{\infty} dx \Psi^\dagger(x,t)
 \left(i \partial_t +\half (\partial_x^2 + x^2 - {q^2 -
 {1\over 4} \over x^2}) + \mu\right) \Psi(x,t) }
In order to express it in terms of the $s$ and $u$ variables we
define new fermionic fields which are related by integral
transforms
 \eqn\psiusda{\eqalign{
 &\Psi_s(s,t)= \int dx \langle s, q| x,q\rangle \Psi(x,t) =
e^{ - i \pi ( { 1 \over 4} - {3 q \over 4}) }
 \int_0^\infty dx \sqrt{2 x s}\,
e^{ { i \over 2} (x^2 + s^2)}
 J_q(\sqrt 2 sx)\Psi(x,t) \cr
 &\Psi_u(u,t) =\int dx \langle u, q| x,q\rangle \Psi(x,t) =
 e^{  i \pi ( { 1 \over 4} - {3 q \over 4}) }
  \int_0^\infty dx \sqrt{2 x u}
   \,
 e^{ -{ i \over 2} (x^2 + u^2)}
 J_q(\sqrt 2 u x)
 \Psi(x,t)\cr
 &\Psi_s(s,t)= \int d u \langle s, q| u, q\rangle \Psi_u(u,t) =
 \int_0^\infty du \sqrt{su}J_q(su)
 \Psi_u(u,t)
 }}
 As in \psir\ it
is convenient to express them in terms of incoming and outgoing
fermions
 \eqn\psiinouta{\eqalign{
 &\Psi^{(in)}(r,t)=e^{r\over 2}\Psi_u(u=e^r)\cr
 &\Psi^{(out)}(r,t)=e^{r\over 2}\Psi_s(s=e^r)\cr
 &\Psi^{(out)}(r,t)= \int_{-\infty}^\infty dr'
 e^{r+r'}J_q(e^{r+r'}) \Psi^{(in)}(r',t)
 }}
It is easy to express the Lagrangian \zala\ in these variables
 \eqn\zalasu{\eqalign{
 {\cal L}&=\int_0^{\infty} ds \Psi_s^\dagger(s,t)\left(i
 \partial_t  +{i \over 2}(s \partial_s +\partial_s s) + \mu
 \right)  \Psi_s (s,t)\cr
 &=\int_0^{\infty} du \Psi_u^\dagger(u,t)\left(i
 \partial_t -{i \over 2} (u \partial_u +\partial_u u) + \mu\right)
 \Psi_u(u,t)\cr
 &=\int_0^{\infty} dr \Psi^{(in)\dagger}(r,t)\left(i
 \partial_t  -i  \partial_r  + \mu \right)  \Psi^{(in)}(r,t)\cr
 &=\int_0^{\infty} dr \Psi^{(out)\dagger}(r,t)\left(i
 \partial_t  + i  \partial_r +  \mu \right)  \Psi^{(out)}(r,t)
 }}
As in the 0B theory we have a $U(1)$ symmetry which rotates $\Psi$
by a phase and corresponds to fermion number conservation.  The
spacetime charge conjugation symmetry, which is  generated by
$(-1)^{F_L}$ on the worldsheet, acts as $q \to -q$. Up to an
overall phase this acts leaving $\Psi^{(in)}$ invariant and
changing $\Psi^{(out)} \to (-1)^q\Psi^{(out)}$, which  is somewhat
trivial. The S-duality transformation,
  $(-1)^{f_L}$,
acts on the parameter $\mu$ as $\mu \to -\mu$, and is a symmetry
only for $\mu=0$. Its action on the fields is
 \eqn\ztwosya{\eqalign{
 &\Psi(x,t) \to   \int_0^\infty dx' \sqrt{xx'} J_q(xx')
 \Psi^\dagger (x',t)\cr
 &\Psi_s(s,t) \to  \Psi_s^\dagger(s,t)\cr
 &\Psi_u(u,t) \to  \Psi_u^\dagger(u,t)\cr
 &\Psi^{(in)}(r,t) \to  \Psi^{(in)\dagger}(r,t)\cr
 &\Psi^{(out)}(r,t) \to \Psi^{(out)\dagger}(r,t)
 }}
It is easy to check that it is a symmetry of the Lagrangian
\zala\zalasu\ with $\mu=0$, and of the transforms \psiusda.  It is
interesting that while in $x$ space the transformation involves an
integral transform, in $s$ and $u$ space the transformation is
local and it is very simple. We conclude that the S-duality
symmetry acts as charge conjugation of  the matrix model fermions
in the $s$, $u$ and $r$ variables.

\subsec{ Computation of the free energies}

We are interested in computing the partition function of the
thermal system. Using the density of states $\rho(\epsilon) = {
\phi'(\epsilon) \over 2 \pi} $ we can write
 the standard expression is
 \eqn\freeeg{\eqalign{
 \log{\cal Z}=&\lim_{\Lambda \to \infty}
 {1\over 2\pi}\int_{-\Lambda}^\infty
 d\epsilon \phi'(\epsilon) \log(1+e^{-2\pi R
 (\epsilon-\mu)})\cr
 =& \lim_{\Lambda\to \infty}
 R\left(-\phi(-\Lambda)(\Lambda+\mu)+ \int_{-\Lambda}^\infty
 d\epsilon
 { \phi(\epsilon)\over 1+e^{2\pi R (\epsilon-\mu)}}\right)
 }}
where $-\Lambda$ is a cutoff on the bottom of the Fermi sea and we
neglected terms which are exponentially small at large $\Lambda$.
In order to simplify the analysis and not worrying about the
$\Lambda$ dependence, we will study the second derivative of
\freeeg
 \eqn\freeegs{ \partial_\mu^2\log{\cal Z}= R \int_{-\infty}^\infty
 d\epsilon \partial_\mu^2{ \phi(\epsilon)\over 1+e^{2\pi R
 (\epsilon-\mu)}} }
which is a convergent integral.

Let us now consider the 0B theory. The determinant of the single
particle S-matrix of the 0B theory is $i e^{i\Phi_B(\epsilon)}$.
It can be expressed as
 \eqn\phizb{\Phi_B(\epsilon)=-i \log\left({\Gamma({1\over 2} -
 i\epsilon) \over \Gamma({1\over 2} + i\epsilon)}\right)=
 \int_0^\infty { d t \over t} \left({\sin(\epsilon t)
 \over  \sinh {t\over 2} }  - 2 \epsilon e^{-t} \right)}
where we have used \formla . Then, \freeegs\ becomes
 \eqn\freeb{\eqalign{
 \partial_\mu^2\log{\cal Z}_B=& R \int_0^\infty { d t \over t}
 \int_{-\infty}^\infty d\epsilon \partial_\mu^2{1\over 1
 +e^{2\pi R (\epsilon-\mu)}} \left({\sin(\epsilon t)
 \over  \sinh {t\over 2}} - 2 \epsilon e^{-t} \right)\cr
 =& - \int_0^\infty { d t \over t} \partial_\mu^2\left(
 { \cos(\mu t)\over 2  \sinh {t\over 2}  \sinh {t\over 2 R}  }
 +R \mu^2 e^{-t} \right)
 }}
We now want to integrate this equation twice with respect to
$\mu$.  Invariance under $\mu \to -\mu$ forbids a term linear in
$\mu$ and the constant term is fixed arbitrarily such that
 \eqn\freebf{\eqalign{
 \log{\cal Z}_B=& - \int_0^\infty { d t \over t}\left(
 { \cos(\mu t)\over 2  \sinh {t\over 2}  \sinh {t\over 2 R}
 }- {2R\over t^2}+\left[{1\over 12} (R+{1\over R})
 +R \mu^2 \right] e^{-t} \right)\cr
 =&\Omega(y=2 i \mu ,R) + \Omega(\bar y=-2 i \mu, R)
 }}

The single particle S-matrix in the 0A theory with nonzero $q$ is
$e^{i\Phi_A(\epsilon)}$.  It can be expressed as
 \eqn\phiza{\Phi_A(\epsilon)= - i \log\left({\Gamma(
 {1\over 2} +{1\over 2}(q - i\epsilon)) \over \Gamma({1\over 2}
 +{1\over 2}(q + i\epsilon))}\right)=
 \int_0^\infty { d t \over t} \left({e^{- q t/2}\sin({\epsilon\
 t\over 2}) \over  \sinh {t\over 2} }  - \epsilon e^{-t} \right)}
where we have used \formla , and we dropped a constant term in the
phase, as well as a term that is linear in $\epsilon$. The term
linear in $\epsilon$ could be removed by doing a rescaling of the
variables $u$ and $s$ that appeared in the 0A discussion\foot{
This term that is linear in $\epsilon$ would have lead to an extra
term proportional to   $  \mu^2 $ in the free energy. We choose to
remove this analytic term by hand.} Then, \freeegs\ becomes
 \eqn\freea{\eqalign{
 \partial_\mu^2\log{\cal Z}_A=&   R \int_0^\infty { d t \over t}
 \int_{-\infty}^\infty d\epsilon \partial_\mu^2{1\over 1
 +e^{2\pi R (\epsilon-\mu)}} \left({e^{- q t/2}\sin({\epsilon\
 t\over 2}) \over  \sinh {t\over 2} }  - \epsilon e^{-t}
 \right)\cr
 =&  - \int_0^\infty { d t \over t} \partial_\mu^2\left({ e^{- q
 t/2} \cos({\mu t\over 2})\over 2  \sinh {t\over 2}  \sinh
 {t\over 4 R}} +{R \mu^2\over 2}  e^{-t} \right)
 \cr
 = & \partial_\mu^2 \left[ \Omega(y=q+ i \mu ,2R) +
 \Omega(\bar y=q- i \mu,2 R) \right]
 }}
Using the definition of the function $\Omega(y,r)$ \ndwdfr\ one
can derive
 \eqn\OmegaT{
 \Omega(yr,{1\over r}) = \Omega(y ,r) -[ {1 \over 24}
 (r + {1\over r}) - {r y^2 \over 8 }] \log r
 }
Using this relation we can check that the 0B answer \freeb\ is the
same as the 0A answer  \freea\ for $q=0$, up to a term involving
$\log r$. This term arises because in 0A and 0B it is natural to
choose the cutoffs to be $R$ independent. On the other hand
T-duality relates them by $\Lambda_{B} = \Lambda_{A} { R_A \over
\sqrt{ 2 \alpha'}} $ (see \tdualityi ). Once we take this into
account,
 the terms that are logarithmic in the cutoff give
a contribution cancelling  the last term in \OmegaT.

\listrefs
\end